\documentclass[3p, times]{elsarticle}
\usepackage{amssymb}
\usepackage{amsmath}
\usepackage{latexsym}
\usepackage{graphicx}
\usepackage{xcolor}
\usepackage{hyperref}
\usepackage{caption}
\usepackage{microtype}
\usepackage{indentfirst}
\usepackage[stable]{footmisc}
\usepackage{listings}
\usepackage{titlesec}
\usepackage{tocloft}
\usepackage{calc}
\usepackage{wrapfig}
\usepackage{subfig}
\usepackage{float}
\usepackage{centernot}
\usepackage{perpage}
\usepackage{mathtools}
\usepackage{stmaryrd}
\usepackage{upquote}
\usepackage{textcomp}

\titleformat{\section}{\large\bfseries}{\thesection}{1em}{}
\titleformat{\subsection}{\large\bfseries}{\thesubsection}{1em}{}
\definecolor{grey}{RGB}{128,128,128}
\definecolor{listings_comment}{RGB}{78,78,78}
\definecolor{listings_base}{RGB}{0,0,0}
\definecolor{listings_caption_font_color}{RGB}{50,50,50}
\definecolor{listings_strings}{RGB}{0,128,0}
\definecolor{listings_keywords}{RGB}{0,0,67}
\definecolor{listings_numbers}{RGB}{0,0,255}
\definecolor{listings_static}{RGB}{102,14,122}
\definecolor{listings_exception}{RGB}{255,10,10}
\definecolor{listings_inline}{RGB}{255,255,255} 
\DeclareCaptionFormat{listing}{{\parbox{\textwidth}{#1#2#3}}}

\newcommand{\static}[1]{{\color{black}\itshape#1}}

\newcommand{\property}[1]{{\color{listings_static}#1}}
\newcommand{\mapitem}[1]{{\color{listings_strings}#1}}
\newlength{\mappingarrowlength}
\newcommand{\mappingarrow}{%
  \underset{\mbox{\footnotesize maps to}}{%
\xmapsto{\hspace*{\mappingarrowlength}}}  
}

\lstdefinelanguage{groovy}[]{java}{
  morekeywords={as,def,in,use,assert,println},
  keywords=[3]{findAll,eachWithIndex,collect,groupBy,inject},  
}

\newcommand\listingtab{.35in}
\lstnewenvironment{msout}{%
\lstset{%
numbers = left,%
numberblanklines = false,%
numberstyle=\color{listings_base},%
numbersep=5pt,%
xleftmargin=.65in,%
extendedchars=true,%
stringstyle={},%
aboveskip=0pt,%
belowskip=0pt,%
framesep=0pt,%
stringstyle={}%
}}{}

\lstnewenvironment{sout}{%
\noindent%
\lstset{%
numbers = left,%
numberblanklines = false,%
numberstyle=\color{listings_base},%
numbersep=5pt,%
xleftmargin=.65in,%
extendedchars=true,%
stringstyle={},%
aboveskip=5pt plus 0pt minus 4pt,%
belowskip=0pt,%
framesep=0pt,%
stringstyle={}%
}}{
\noindent\hspace{\listingtab}\rule{\textwidth-\listingtab-\listingtab}{.1pt}%
\vspace{3pt}%
\newline%
}

\lstnewenvironment{soutnoframe}{%
\lstset{%
numbers = left,%
numberblanklines = false,%
numberstyle=\color{listings_base},%
numbersep=5pt,%
xleftmargin=.65in,%
extendedchars=true,%
stringstyle={},%
aboveskip=1pt plus 0pt minus 4pt,%
belowskip=9pt,%
framesep=0pt,%
stringstyle={}%
}}{}

\newcommand\inlinecode[1]{%
  \begingroup%
  \lstset{deleteemph=[1]{%
      Expand, ExpandAll, ExpandDenominator, ExpandNumerator, %
      EliminateMetrics, Differentiate, Factor, ExpandAndEliminate, %
      Conjugate, DiracTrace,  Matrix1, UnitaryTrace,%
      EliminateDueSymmetries, Reverse,%
      Together, TogetherFactor, LeviCivitaSimplify,%
      LatinLower, LatinUpper, Matrix1, Matrix2%
     },%
    deleteemph=[2]{%
      indices, free, inverted,%
      upper, lower},%
    deleteemph=[3]{%
       Redberry, LaTeX, UTF8, WolframMathematica%
    },%
    stringstyle={},%
    keywordstyle={},%
    basicstyle=\small\ttfamily\color{listings_base},%
 }%
  \lstinline!#1!%
  \endgroup%
}

\DeclareCaptionFont{black}{\color{black}}
\DeclareCaptionFont{white}{\color{white}}
\DeclareCaptionFont{listings_caption_font}{
\color{listings_caption_font_color}\small\bf}

\definecolor{urlcolor}{RGB}{15,25,112}
\hypersetup{
    bookmarks=true,         
    unicode=false,          
    pdftoolbar=true,        
    pdfmenubar=true,        
    pdffitwindow=false,     
    pdfstartview={FitH},    
    pdftitle={Redberry},    
    pdfauthor= {Stanislav Poslavsky},     
    pdfsubject={Symbolic tensor computations}, 
    pdfcreator={Stanislav Poslavsky},
    pdfproducer={hz}, 
    pdfkeywords={symbolic tensor computations tensor computer algebra system}, %
    pdfnewwindow=true,      
    colorlinks=true,       
    linkcolor=black,          
    citecolor=black,        
    filecolor=magenta,      
    urlcolor=urlcolor           
}

\usepackage{lipsum}
\makeatletter
\def\ps@pprintTitle{%
 \let\@oddhead\@empty
 \let\@evenhead\@empty
 \def\@oddfoot{}%
 \let\@evenfoot\@oddfoot}
\makeatother
\begin{document}
\begin{frontmatter}
\title{Introduction  to  Redberry: a computer algebra system designed for tensor 
manipulation}

\author[a]{Dmitry Bolotin\corref{DB}}
\author[b]{Stanislav Poslavsky\corref{SP}}

\cortext[DB] {\textit{E-mail address:} 
\href{mailto:bolotin.dmitriy@gmail.com}{bolotin.dmitriy@gmail.com}}
\cortext[SP] {\textit{E-mail address:} 
\href{mailto:stvlpos@mail.ru}{stvlpos@mail.ru}}
\address[a]{Institute of Bioorganic Chemistry of the Russian Academy  of Sciences, 
117997 Moscow, Russia}
\address[b]{Institute for High Energy Physics, 142281 Protvino, Moscow Region, 
Russia}

\begin{abstract}
In this paper we introduce Redberry --- an open source computer algebra system with native support of tensorial expressions. It provides basic computer algebra tools (algebraic manipulations, substitutions, basic simplifications etc.) which are aware of specific features of indexed expressions: contractions of indices, permutational symmetries, multiple index types etc. Redberry supports conventional \LaTeX-style input notation for tensorial expressions. The high energy physics package includes tools for Feynman diagrams calculation: Dirac and SU(N) algebra, Levi-Civita simplifications and tools for one-loop calculations in quantum field theory. In the paper we give detailed overview of Redberry features: from basic manipulations with tensors to real Feynman diagrams calculation, accompanied by many examples. Redberry is written in Java 7 and provides convenient Groovy-based user interface inside the high-level general purpose programming language environment.
\end{abstract}


\end{frontmatter}

\noindent {\bf PROGRAM SUMMARY}

\vspace{8pt}

\begin{small}
\noindent
{\em Authors:}  Dmitry Bolotin and Stanislav Poslavsky\\
{\em Program Title:} Redberry\\
{\em Licensing provisions:} GNU General Public License, version 3\\
{\em Programming language:} Java, Groovy\\
{\em Computer:} All\\
{\em Operating system:} Any platform supporting Java 7 and higher\\
{\em RAM: } Problem dependent\\
{\em No. of lines in distributed program, including test data, etc.:} 130 407\\
{\em No. of bytes in distributed program, including test data, etc.:} 5 744 913
{\em External routines/libraries:} Apache Commons Math [1], Trove4j [2], Junit [3]\\
{\em Subprograms used:} Java Algebra System [4]\\
{\em Nature of problem:} Algebraic manipulation with tensorial objects\\
{\em Additional comments:} The latest updates can be found at \href{http://redberry.cc}{http://redberry.cc}. Installation instructions, comprehensive documentation, VCS and issue tracker can be also found at Redberry website.

\vspace{6pt}

\noindent{\em References:}

\noindent [1] Apache Commons Math, \href{http://commons.apache.org/proper/commons-math/}{http://commons.apache.org/proper/commons-math/} 

\noindent [2] Trove4j, \href{http://trove.starlight-systems.com/}{http://trove.starlight-systems.com/} 

\noindent [3] Junit, \href{http://junit.org/}{http://junit.org/}

\noindent [4] Java Algebra System, \href{http://krum.rz.uni-mannheim.de/jas/}{http://krum.rz.uni-mannheim.de/jas/} 
\end{small}
\vspace{9pt}
\hrule

\newpage
\begingroup
\hrule
\hypersetup{ linkcolor=black }
\renewcommand{\contentsname}{{\normalsize \bf CONTENT}}
\topskip0pt
\tableofcontents
\vspace{12pt}
\hrule
\endgroup

\section{Introduction}
\subsection{Background}
General-purpose computer algebra systems (CASs) have become an essential part of many scientific calculations. Focusing on the area of theoretical physics and particularly high energy physics, one can note that there is a wide area of problems that deal with tensors (or more generally --- objects with indices). Specifically in this work we focus on the algebraic manipulations with abstract indexed expressions that forms a substantial part of computer aided calculations with tensors in this field of science. Today, there are many packages both on top of general-purpose systems (xAct \cite{xAct}, Tensorial \cite{Tensorial}, Ricci \cite{Ricci}, Maple Physics \cite{MaplePhysics} etc.) and standalone tools (Cadabra \cite{Cadabra,Cadabra2}, SymPy \cite{SymPy}, GiNaC \cite{GiNaC1,GiNaC2} etc.) that cover different topics in symbolic tensor calculus (some comparison of existing systems can be found in Sec.~\ref{sec_comparison}). However, it cannot be said that current demand on such a software is fully satisfied \cite{RUDN}.

The main distinguishing feature of tensorial expressions (in comparison with ordinary indexless) is that Einstein notation brings an additional structure to the mathematical expressions; specifically, contractions between indices forms a mathematical graph (which sometimes expressed using Penrose graphical notation). This additional structure must be reflected in the computer representation of indexed object, which makes the existing pure list-based CASs not a good basis for implementing algorithms for tensor manipulation \cite{Cadabra2}. Because of this structure, a computer implementation even of such a fundamental atomic operation like expression comparison becomes much more complicated (for details see Sec.~\ref{sec_IndexMapping}), which in turn complicates implementation of general operations such as substitutions, reducing similar terms etc. Since any routine manipulation comprise a long sequence of these general operations, their performance is critical for real-world problems.

Here we present a new computer algebra system Redberry aimed on algebraic manipulations with tensorial expressions. Albeit the architecture of Redberry is not focused on solving particular problems in high energy physics, it contains a number of high-level features for calculations in the area, since it is the main field of author's scientific interests. Because of this also, majority of the examples in the paper will be focused on quantum field theory computations.

The key features of Redberry include: native dummy indices handling (including automatic clash resolution), support of arbitrary permutational symmetries of tensor indices, extensive tools for comparison of tensorial expressions, multiple index types, \LaTeX-style input/output and a comprehensive set of tensor-specific transformations. Out of the box Redberry provides a package for computations in high energy physics including tools for Feynman diagrams calculation (Dirac \& SU(N) traces, simplification of Levi-Civita tensors etc.) and for calculation of one-loop counterterms in general field theory.

Redberry core is written in Java, while the user interface is written in Groovy and is intended to be used within the Groovy environment; in this manner Redberry functionality is available through the modern high-level general-purpose programming language, allowing to use both general programming capabilities and tensor computer algebra in the same environment. As an example of successful employment of such a model, one can mention SymPy \cite{SymPy} computer algebra system, which uses Python programming language as a main user interface. Despite Redberry does not provide GUI by itself, the modern open source IDEs\footnote{we recommend JetBrains \href{http://www.jetbrains.com/idea/}{IntelliJIDEA} because of its great support of Groovy} give a very convenient way to work with Redberry including syntax highlighting and code completion.

\subsection{The introductory example}
\label{sec_IntroductoryExample}
Let's start from the minimal real example to quickly dive into Redberry (the next sections cover every aspect in details). The example demonstrates calculation of differential cross section of the Compton scattering in scalar massless Quantum Electrodynamics (the reader which is not familiar with this area can skip physical details and focus just on the algebraic operations):
\begin{lstlisting}[label=lst:ComptonInScalarQED,%
frame=tb,%
numbers=left]
//photon-scalar-scalar vertex
def V1 = 'V_i[p_a, q_b] = -I*e*(p_i + q_i)'.@\ttt@
//photon-photon-scalar-scalar vertex
def V2 = 'V_{ij} = 2*I*e**2*g_{ij}'.@\ttt@
//scalar propagator for massless particle
def P = 'D[k_a] = -I/(k^a*k_a)'.@\ttt@
//matrix element
def M = '''M^ij = V^i[k3_a, k3_a + k1_a]*D[k3_a + k1_a]*V^j[-k4_a, -k3_a - k1_a] 
                  + V^j[k3_a, k3_a - k2_a]*D[k3_a - k2_a]*V^i[-k3_a + k2_a, -k4_a]
                  + V^ij'''.@\ttt@
//substituting vertices and propagator in matrix element
M = (V1 & V2 & P) >> M
//squared matrix element (minus is due to complex conjugation)
def M2 = M >> 'M2 = -M_ij*M^ij'.@\ttt@
//expand squared matrix element and eliminate metrics and Kronecker deltas
M2 = (ExpandAll & EliminateMetrics & 'd^i_i = 4'.@\ttt@) >> M2
//massless particles: k1_a*k1^a = 0, k2_a*k2^a = 0 etc.
for (def i in 1..4)
    M2 = "k@\$\{i\}@_a*k@\$\{i\}@^a = 0".@\ttt@ >> M2
//momentum conservation
M2 = ('k1_a*k2^a = k3_a*k4^a'.@\ttt@ & 'k1_a*k3^a = k2_a*k4^a'.@\ttt@ & 'k1_a*k4^a = k2_a*k3^a'.@\ttt@) >> M2
//factor terms
M2 = Factor >> M2
println M2
\end{lstlisting}
This script will print a well-known expression for the squared matrix element of the Compton scattering in massless scalar QED:
\begin{equation*}
 |\mathcal{M}|^2  = -\frac{e^{4}}{2} \frac{-18 (k_2 k_3) (k_2 k_4)+(k_3 
k_4)^{2}+(k_2 k_3)^{2}+2( (k_2 k_3)- (k_2 k_4)) (k_3 k_4)+(k_2 k_4)^{2}}{(k_2 
k_3)(k_2 k_4)}, 
\end{equation*}
where $(k\,p)$ denotes scalar product $k_i p^i$.

The code above speaks for itself, so leaving aside the physical aspects of the problem, let's focus only on programming aspects. \inlinecode{'...'.t} construct converts a string representation into a computer object. The input notation for tensors is the same as used in \LaTeX{} with minor syntax modifications\footnote{one can omit curly braces where it does not cause ambiguity.}. Redberry uses the predefined notation for some built-in tensors, such as \inlinecode{I} for imaginary one, or \inlinecode{g_ij} and \inlinecode{d^i_j} for metric tensor and Kronecker delta respectively. Thus, for example, when we perform contractions with metrics and deltas in the line 16, Redberry automatically takes into account that $g_{ai}\,g^{aj} = \delta_i{}^j$. However, since the dimension of the space affects Kronecker trace value and it is not specified anywhere explicitly, we substitute the trace manually in line 16.

As could be seen from the example, state of index (upper or lower, i.e. contravariant or covariant) is important --- indices are considered to be contracted if they have different states.

All transformations in Redberry (e.g. substitutions, \inlinecode{Expand}, \inlinecode{EliminateMetrics} etc.) are first-class objects and can be assigned to variables. They can be applied to mathematical expression using \inlinecode{>>} operator. Detailed discussion on Redberry transformations usage and list of basic built-in transformations can be found in Sec.~\ref{sec_BasicTransformations}.

The last point that should be discussed here is looping construct used in lines 18, 19.  Here standard Groovy syntax used to apply several similar substitutions ($k_{1i} k_1{}^i = 0$, $k_{2i} k_2{}^i = 0$ and so on).

\subsection{Paper structure}
The whole paper is accompanied by many Groovy examples. Each example can be executed by simple wrapping the code in the following way:
\begin{lstlisting}[frame=tb,escapechar=!]
@Grab(group = 'cc.redberry', module = 'groovy', version = '1.1.8')
import cc.redberry.groovy.!Redberry!

import static cc.redberry.core.tensor.Tensors.*
import static cc.redberry.groovy.RedberryPhysics.*
import static cc.redberry.groovy.RedberryStatic.*

use(!Redberry!){  
  //example code  
}
\end{lstlisting}
Redberry and all required dependencies will be downloaded automatically. Some examples also may require additional \inlinecode{import} statements, in this case it will be noted in a footnotes.

The remainder of the paper is organized as follows. In Sec.~\ref{sec_Basics} we give description of Redberry basic functionality, needed for usage and understanding of main Redberry principals. In Sec.~\ref{sec_BasicTransformations} one can find a list of selected transformations and general aspects concerning their application. Sec.~\ref{sec_PhysExamples} illustrates Redberry application in high energy physics: calculation of Compton scattering in QED and calculation of one-loop counterterms of vector field operator in a curved space-time. In Sec.~\ref{sec_InternalArchitecture} we consider selected architectural solutions used in Redberry. Finally in Sec.~\ref{sec_comparison} we compare Redberry functionality and performance with selected existing systems.

\section{Basics}
\label{sec_Basics}
\subsection{Tensors and indices}
\label{sec_architecture}
Redberry is written in Java and makes extensive use of its object-oriented features. Though user may not be familiar with object oriented programming to use Redberry, it is still useful to understand primitive object types that are used in the CAS. There are three central object types in Redberry: \inlinecode{Tensor}, \inlinecode{Indices} and \inlinecode{Transformation}. Objects of the same type share common properties and can be manipulated in a common way. Each of these types may have several inheritors like \inlinecode{Sum}, \inlinecode{Product} and \inlinecode{SimpleTensor} (for simple tensor like $x_m$) for \inlinecode{Tensor}, or \inlinecode{Expand} for \inlinecode{Transformation}. Tensors and their indices are considered in this section, while transformations are discussed in Sec.~\ref{sec_BasicTransformations}.

Each mathematical expression in Redberry is a \inlinecode{Tensor}. Any \inlinecode{Tensor} has \inlinecode{Indices} and content (summands in case of sum, arguments in case of functions, etc.; thus, tensors in Redberry are containers of other tensors). Here is how these properties can be accessed via Groovy syntax:
\begin{lstlisting}[frame = t,numbers=left,name=Primiteves]
def p = 'F^{A}_{B \\mu \\nu} * a'.@\ttt@ 
println p.@\property{class}@.@\property{simpleName}@ //name of the Tensor subtype
\end{lstlisting}
\begin{msout}
Product
\end{msout}
\begin{lstlisting}[numbers=left,name=Primiteves]
println( [p[0], p[1]]) //getting element by index
\end{lstlisting}
\begin{msout}
[a, F^{A}_{B \\mu \\nu}]
\end{msout}
\begin{lstlisting}[numbers=left,name=Primiteves]
p.each{ i -> println i } //enumerating elements
\end{lstlisting}
\begin{msout}
a
F^{A}_{B \\mu \\nu}
\end{msout}
\begin{lstlisting}[numbers=left,name=Primiteves]
println p.size() //size of tensor as container
\end{lstlisting}
\begin{msout}
2
\end{msout}
\begin{lstlisting}[numbers=left,name=Primiteves]
println p.@\property{indices}@ //accessing indices of tensor
\end{lstlisting}
\begin{sout}
^{A}_{B \\mu \\nu}
\end{sout}
The meaning of the above lines of code is pretty evident. In order to create tensors programmatically, Redberry defines all arithmetic operations for  \inlinecode{Tensor} objects:
\begin{lstlisting}[frame=t]
def x = 'x'.@\ttt@, y = 'y'.@\ttt@
def t = (-x + sin(y))**2 / (x + sin(-y)) + x - sin(y)
println t
\end{lstlisting}
\begin{sout}
2*Sin[-y]+2*x
\end{sout}
However, such syntax may be inconvenient when dealing with indexed objects, since names of variables do not reflect structure of indices  of expressions.

Tensors are immutable and all modification operations return new instances:
\begin{lstlisting}[frame=t]
def t = 'A_i + C_i'.@\ttt@
def x = t.set(0, 'B_i'.@\ttt@), y = t.remove(0)
println( [t, x, y] )
\end{lstlisting}
\begin{sout}
[A_i+C_i, B_i+C_i, C_i]
\end{sout}

Presence of indices of expressions is a main distinguishing property of tensor-oriented CASs. As mentioned above, there is an \inlinecode{indices} property defined for each expression in Redberry. Returned object is an object of type \inlinecode{Indices} and have a number of methods and properties to work with it. Here are some examples of possible operations with indices object: 
\begin{lstlisting}[name = ExIndices, frame = t,%
  numbers = left,
  label = lst:ExIndices
]
def t = '2*x_am*f^m*(a^n + b^n)'.@\ttt@
def ind = t.indices
println( [ind, ind.size()] )
\end{lstlisting}
\begin{msout}
[^{mn}_{am}, 4]
\end{msout}
\begin{lstlisting}[name = ExIndices, numbers = left]
println ind.free
\end{lstlisting}
\begin{msout}
^{n}_{a}
\end{msout}
\begin{lstlisting}[name = ExIndices, numbers = left]
println ind.inverted
\end{lstlisting}
\begin{msout}
^{am}_{mn}
\end{msout}
\begin{lstlisting}[name = ExIndices, numbers = left]
println( [ind.upper, ind.lower] )
\end{lstlisting}
\begin{sout}
[^{mn}, _{am}]
\end{sout}
Methods used in the above example, are inherent in any indices object. Their names clearly implies their meaning.

Although the presence of indices is inherent in all tensors, different types  of expressions have different subtypes of \inlinecode{Indices} objects. This difference arises from availability or unavailability of information about indices ordering. Consider indices of simple tensor:
\begin{lstlisting}[frame = t]
def simple = 'F_{mn}^{\\beta\\alpha}_{ba\\alpha}'.@\ttt@
println simple.indices
\end{lstlisting}
\begin{sout}
_{mnba}^{\beta\alpha}_{\alpha}
\end{sout}
We shall call indices of simple tensors as "simple indices" (subtype \inlinecode{SimpleIndices}). For "simple indices" the order of indices is defined, which is its main distinguishing property. In other words, permutation of indices will result in changing of mathematical sense of the particular simple tensor (unless this tensor is symmetric with respect to this permutation, see Sec.~\ref{sec_Symmetries}). However, as we previously mentioned, each index belongs to some index type (e.g. Greek upper/lower case or Latin upper/lower case, etc.). Indices of different types are considered to have different mathematical nature (e.g. Greek indices are Lorentz, Latin are SU(N) etc.), so the relative positions of indices with different types is not important. Thus, Redberry sorts indices of simple tensors according to their types, preserving the relative order of indices belonging to the same type (see the above example). Indices of \inlinecode{TensorField} (a subtype of \inlinecode{Tensor} which denotes tensor function like $f_\mu (x_\nu, \, y_\alpha + z_\alpha)$ ) are also "simple indices"and follows the same ordering rules.

Another type of indices is inherent in all other types of tensors. Consider the following product:
\begin{lstlisting}[frame = t]
def pr = 'F_{mn}*F^{\\beta\\alpha}*F_{ba\\alpha}'.@\ttt@
println pr.indices
\end{lstlisting}
\begin{sout}
^{\alpha\beta}_{abmn\alpha}
\end{sout}
From the mathematical point of view, order of product indices is undefined. This allows to set up some certain ordering rules (mainly for technical reasons, related to performance). As one can see from the example, all indices are sorted according to the following rules: first upper then lower, first Latin then Greek, in each group of indices with the same type indices are sorted in lexical order. The similar rules are adopted for sums:
\begin{lstlisting}[frame = t]
def sum = 'R^a_amn^\alpha+K^i_inm^\alpha'.@\ttt@
println sum.indices
\end{lstlisting}
\begin{sout}
^{\alpha}_{mn}
\end{sout}
The only difference is that according to the sense of sum, its indices are actually free indices of its summands.

All methods from above listings (\inlinecode{.inverted}, \inlinecode{.free} etc.), return objects with the same rules of ordering as in the initial indices object.

In conclusion, it should be noted that at the low-level, Redberry stores each single index as 32-bit integer that encodes all information about index: state, type and name (serial number in the alphabet). Within a particular index type, there are $2^{16}$ possible names, which can be inputted using the subscript like
\begin{lstlisting}[frame = tb]
def t = 'X_{a_1 a_{122} b_9 \\mu_3}'.@\ttt@
\end{lstlisting}
There are many utility methods to operate with single index, so user is freed from knowing the particular bits layout.Further documentation on tensors, indices and single index manipulation can be found on \href{http://redberry.cc/documentation}{Redberry website}.

\subsection{Einstein notation}
At this point the basic conventions arising from Einstein notation should be clarified. As was previously mentioned, Redberry distinguishes covariant (lower) and contravariant (upper) indices. The corresponding property of single index we call {\itshape state}. Two indices are considered to be contracted if and only if they have similar names and types but different states. As a consequence of this convention there are some natural restrictions on general structure of the expressions.

First of all, the following notation (used, for example, in Euclidean space) is illegal in Redberry:
\begin{lstlisting}[frame = t]
def t = 'F_aa'.@\ttt@ //correct input is F_a^a or F^a_a
\end{lstlisting}
\begin{sout}
@\color{listings_exception}InconsistentIndicesException@
\end{sout}
The error also occurs when expression is meaningless because of several
same indices coexist in a product:
\begin{lstlisting}[frame = t]
def t = 'F_ij*M^i*N^j*K^j'.@\ttt@ //meaningless expression
\end{lstlisting}
\begin{sout}
@\color{listings_exception}InconsistentIndicesException@
\end{sout}
The entire architecture of Redberry built in such a way that the above illegal situations can not ever arise during manipulations and user do not need to take care about it.

Other thing arises with dummy indices in products, where one or more multipliers are sums. It is convenient to write
\[
 F_{\mu\nu} (A^{\alpha\beta} + M_\mu N^{\mu\alpha\beta}).
\]
There is no problem here, since dummy index $\mu$ is in the scope of sum, which "outer"\footnote{If consider sum as a one tensor $T^{\alpha\beta}=A^{\alpha\beta} + M_\mu N^{\mu\alpha\beta}$} indices are upper ${\alpha\beta}$, while the first multiplier indices are lower ${\mu\nu}$. However, if we try to expand brackets in this expression naively (i.e. without relabelling of dummy index $\mu$) we shall face the ambiguity as described in the previous examples. Thus it is better to relabel such dummy indices in sums right after parsing, and this is what Redberry does:
\begin{lstlisting}[frame = t]
def t = 'F_mn*(A^ab + M_m*N^mab)'.@\ttt@
println t
\end{lstlisting}
\begin{sout}
F_mn*(A^ab + M_c*N^cab)
\end{sout}
Here we see, that conflicting dummy index \inlinecode{m} was automatically renamed to \inlinecode{c}.

\subsection{Standard form of mathematical expressions}
\label{sec_StandardForm}
A core feature of any CAS is its ability to reduce arbitrary expression to some standard form (SF) in which it is then used everywhere in manipulations. This approach facilitates comparison and matching of expressions and gives a way for more robust and fast algorithms for almost all CAS operations. Redberry uses the same paradigm: any intermediate and resulting expression is guaranteed to be in the SF.

Consider the following examples, which give an idea of SF in Redberry:
\begin{lstlisting}[name = StandardFormSymbolic,  frame = t]
def a = '(a-b)+c+(b-a)'.@\ttt@,
     b = 'c*(a-b)*(b-a)/c'.@\ttt@,
     c = '2/3-27**(1/3)/9'.@\ttt@,
     d = 'Sin[2 + 2.*I]**(1/4)'.@\ttt@
println( [a, b, c, d] )
\end{lstlisting}
\begin{sout}
[c, -(a-b)**2, 1/3, 1.38307 - 0.144188*I]
\end{sout}
The first example demonstrates that Redberry performs the reduction of similar terms. The second --- that same (to within a sign) multipliers are collected into powers or reduced. Numbers are always collected and reduced if possible. If expression contains floating-point numbers, then it will be completely reduced (calculated). This behaviour is similar to the majority of symbol-oriented computer algebra systems and needs no more detailed explanation.

The additional conventions on standard form arise in expressions that contain tensors. The most remarkable convention is on the SF of sum. It is best to demonstrate it by example:
\begin{lstlisting}[name = SumConvention, frame = t]]
println 'a*F_mn + (a + b)*F_mn'.@\ttt@
\end{lstlisting}
\begin{msout}
(2*a+b)*F_{mn}
\end{msout}
\begin{lstlisting}[name = SumConvention]
println '(x_a^a + y_b^b)*X_m*X^m + (z_n^n - y_d^d)*X_a*X^a'.@\ttt@
\end{lstlisting}
\begin{sout}
(x_{a}^{a}+z_{n}^{n})*X^{m}*X_{m}
\end{sout}
As one can see, Redberry tries to factor out parts of products which contain all multipliers which have nonzero number of ``outer'' indices\footnote{multipliers like $(\kappa{}^\mu{}_\mu+\sigma{}^\alpha{}_\alpha)$ are considered to have no ``outer'' indices, wile tensors like $x{}^m{}_m$ are considered to have ``outer'' indices, because they contribute to a whole product indices.}.

The other important detail is ordering of summands and multipliers within sums and products. Elements of sums and products are sorted by their 32 bit hash codes. Hashes for simple tensors (e.g. \inlinecode{x} or \inlinecode{k_p}) are generated randomly at each Redberry run, while hashes of complicated tensors (e.g. \inlinecode{Sin[x]*k_i}) are calculated  according to certain complex rules\footnote{The most important property of hash functions defined for complex tensors is its "insensibility" for particular names of indices but "sensibility"for their contractions. So, renaming of dummy or free indices does not affect hash code, but any modification of structure of contractions (e.g. contraction of two free indices) does.}. Usage of the pseudorandom generator allows to obtain nearly uniform distribution of hashes of tensors, which significantly improves performance. However, the ordering of expressions changes from run to run, and e.g. product \inlinecode{a*b*c} will be sorted differently at different runs (\inlinecode{b*c*a} or \inlinecode{b*a*c} etc.). Of course, all similar expressions will have similar ordering in current session. Still, Redberry have tools to fix the seed of pseudorandom generator inside, so that expressions will have same ordering from run to run.

It should be clarified that reduction to SF is a light-weight operation which does not perform any time-consuming simplifications. 

In contrast to many other tensors-oriented CASs, like xAct \cite{xAct} or Cadabra \cite{Cadabra}, Redberry does not use so-called indices canonicalization (sorting) approach. It uses another approach for the problem of tensors comparison and simplification, which will be discussed in more detail in Sec.~\ref{sec_IndexMapping}~and~\ref{sec_versus}.

In conclusion, authors want to emphasize that user can rely on the fact that inside any calculation all expressions are reduced to the SF. This fairly simplifies implementation of custom transformations and algorithms.

\subsection{Symmetries of tensors}
\label{sec_Symmetries}
The next distinctive feature of tensors is their symmetries. Consider symmetries under permutations of indices. Permutational symmetries in Redberry can be defined for indices of simple tensors and tensor fields. For example, permutational symmetries of Riemann tensor can be defined in the following way:
\begin{lstlisting}[frame = tb]
@\addSymmetry@ 'R_abcd',  [[0, 2], [1, 3]].@\ppp@ //permutation in cycle notation
@\addSymmetry@ 'R_abcd', -[1, 0, 2, 3].@\ppp@ //permutation in one-line notation
\end{lstlisting}
Method \inlinecode{addSymmetry} have two arguments: a simple tensor (or its string representation) and a permutation. Redberry has internal representation of permutations\footnote{For a comprehensive documentation on permutations and permutation groups in Redberry, see \href{http://redberry.cc/documentation:guide:permutations_and_permutation_groups}{online documentation}.} and in order to convert array to permutation one can use \inlinecode{.p} property followed after the array written in one-line or disjoint cycles notation. Minus (used in the last line) converts symmetry to antisymmetry and vice versa.

Once set, symmetries of tensor affects all further manipulations with it. For example, if Riemann symmetries are set up, then the following code automatically gives zero:
\begin{lstlisting}[frame = t]
println 'R^abcd*R_efdc*R^ef_ab + R_rc^df*R_ab^rc*R_fd^ba'.@\ttt@
\end{lstlisting}
\begin{sout}
0
\end{sout}
Here zero was returned right after parsing; this is because Redberry automatically reduced sum to the standard form and 
\[
R^{abcd} R_{efdc} R^{ef}{}_{ab} = -R_{rc}{}^{df} R_{ab}{}^{rc} R_{fd}{}^{ba}
\]
according to the specified symmetries. Such architecture requires user to set all symmetries of simple tensor before it will be parsed inside any complicated structure like sum or product. If one try to add symmetry to tensor which is already in use in some complicated expression, then the exception will be thrown.

Internally, Redberry aggregates symmetries of simple tensor in a special container which can be accessed using \inlinecode{.symmetries} property of tensor indices. When all generators are specified, Redberry uses a \inlinecode{PermutationGroup}\footnote{The internal implementation is based on base and strong generating set of groups (using Schreier-Sims algorithm). Redberry provides methods for coset enumeration, searching for centralizers, stabilizers, etc. For details see \href{http://redberry.cc/documentation:ref:permutationgroup}{online documentation}.} to hold and manipulate symmetries:
\begin{lstlisting}[frame = t]
def t = 'R_abcd'.@\ttt@ //assume Riemann symmetries are already defined
def symmetries = t.@\property{indices}@.@\property{symmetries}@
def group = symmetries.@\property{permutationGroup}@
println group.order() //total number of all permutations
\end{lstlisting}
\begin{msout}
8
\end{msout}
\begin{lstlisting}
println  group.setwiseStabilizer(2, 3) //compute setwise stabilizer of set [2, 3]
\end{lstlisting}
\begin{msout}
Group( -[[0, 1]], -[[2, 3]] )
\end{msout}
\begin{lstlisting}
//define some other permutation group
def oth = @\static{Group}@(-[[0, 2, 1, 3]], -[[0, 1]], [[2, 3, 4, 5, 6]])
println group.intersection(oth) //compute intersections of groups
\end{lstlisting}
\begin{sout}
Group( -[[2, 3]], +[[0, 2], [1, 3]] )
\end{sout}

One should be careful when attaching antisymmetries. Consider the following code:
\begin{lstlisting}[frame = t]
def t = 'R_abcd'.t
@\static{addSymmetries}@ t, [2, 3, 0, 1].@\ppp@, -[1, 0, 2, 3].@\ppp@, @
@-[3, 2, 1, 0].@\ppp@
def gr =  @
@t.@\property{indices}@.@\property{symmetries}@.@\property{permutationGroup}@
println gr.order()
\end{lstlisting}
\begin{sout}
@\color{listings_exception}InconsistentGeneratorsException@
\end{sout}
Exception is thrown on the last line since that the last attached permutation is a combination of two previous, but its sign is different. Thus, some combinations of symmetries and antisymmetries can come into conflict, which causes an error in Redberry.

Redberry provides tools to find permutational symmetries of complicated tensors. Consider the following example:
\begin{lstlisting}[frame = t, numbers = left]
@\addSymmetry@ 'R_abc', -[1, 0, 2].@\ppp@
@\addSymmetry@ 'A_ab', [1, 0].@\ppp@
def t = '(R_abc*A_de + R_bde*A_ac)*A^ce + R_adb'.@\ttt@
def symmetries = @\static{findIndicesSymmetries}@('_abd'.@\property{si}@, t)
for (s in symmetries)
    println s
\end{lstlisting}
\begin{sout}
+[]
-[[0, 2]]
\end{sout}
The first permutation is identity, while the second is nontrivial\footnote{Method used in the fourth line takes simple indices as the first argument in order to define the relative order of indices in tensor. \inlinecode{'...'.si} construction is used to parse simple indices}.

While the permutational symmetries of tensors are well covered in a number of tensors-oriented CASs, the so-called multi-terms symmetries (like Bianchi identities) are faintly covered or absent at all in the majority of existing systems. In fact, authors know only one system --- Cadabra \cite{Cadabra}, which fully supports multi-term symmetries. The basic idea utilized by Cadabra system is usage of Young tableau projectors to reduce expressions to the simplified form. Consider the following identity\footnote{This example is taken from Section~2.2 of the {\itshape 'Cadabra: reference guide and tutorial'} by Kasper Peeters, which is available on Cadabra \href{http://cadabra.phi-sci.com/cadabra.pdf}{web site}.}:
\begin{equation}
\label{eq:WeylIdentity}
W_{u}{}^{vs}{}_{w} W_{tv}{}^{qw} 
W_{p}{}^{t}{}_{r}{}^{u} W^{p}{}_{q}{}^{r}{}_{s}
-
W^{sv}{}_{u}{}^{w} W_{rvtw} 
W_{p}{}^{qtu} W^{p}{}_{q}{}^{r}{}_{s}
= 
W_{s}{}^{pd}{}_{a} W^{ms}{}_{cd} 
W^{n}{}_{pb}{}^{c} W_{mn}{}^{ab}
-
\frac{1}{4} W^{n}{}_{s}{}^{d}{}_{c} W^{m}{}_{p}{}^{c}{}_{d} 
W^{ps}{}_{ba} W_{mn}{}^{ab}
\end{equation}
where $W_{abcd}$ is a Weyl tensor. In order to proof this identity using a Young
projector, one need to apply the following substitution to the above expression:
\begin{equation}
\label{eq:WeylYoung}
W_{abcd} = \frac{1}{3} \, (2\,W_{abcd}-W_{adbc}+W_{acbd})
\end{equation}
This identity is derived from the Ricci cyclic identity and ordinary permutational 
symmetries of the Weyl tensor. The following Redberry code proofs the identity 
(\ref{eq:WeylIdentity}):
\begin{lstlisting}[
frame = t,%
numbers =left%
]
@\static{addSymmetries}@ 'W_abcd', [[0, 2], [1, 3]].@\ppp@, -[1, 0, 2, 3].@\ppp@
def t = ('W^p_q^r_s*W_p^t_r^u*W_tv^qw*W_u^vs_w'
        + ' - W^p_q^r_s*W_p^qtu*W_rvtw*W^sv_u^w'
        + ' - W_mn^ab*W^n_pb^c*W^ms_cd*W_s^pd_a'
        + ' + 1/4*W_mn^ab*W^ps_ba*W^m_p^c_d*W^n_s^d_c').@\ttt@
def s = 'W_mnpq = 1/3*(2*W_mnpq - W_mqnp + W_mpnq)'.@\ttt@
def r = (s & Expand) >> t
println r
\end{lstlisting}
\begin{sout}
0
\end{sout}
At the moment Redberry have no built-in functionality to construct substitutions like (\ref{eq:WeylYoung}) based on the Young projectors. But, as could be seen from the above example, if one define the corresponding substitution manually, then Redberry allows to work with multi-term symmetries in the Cadabra way. Authors plan to incorporate support of Young projectors in the upcoming releases of Redberry.

\subsection{Derivatives}
\label{sec_Derivatives}
Representation  of derivatives in Redberry is very similar to other systems and is very close to standard mathematical sense of this concept. However, presence of indices brings new features of derivatives with respect to indexed arguments.

Consider first the common mathematical notation for derivatives of ordinary functions. The standard notation $f'(y)$ is really a shorthand for $\left. \frac{d}{d x} f(x)\right|_{x = y}$ etc. In Redberry we use the {\ttfamily'$\sim$'} symbol followed by the derivative order (or orders in case of function with several arguments) instead of primes, so, for example, the following expression
\begin{lstlisting}
def t = 'f@\listingtildestr@(3)[x**2]'.@\ttt@
\end{lstlisting}
represents \(\displaystyle \left. \frac{d^3}{d t^3} f\left(t\right) \right|_{t = 
x^2} \). When substituting e.g. \(f(t) = \sin t\) in the above expression one will 
have
\begin{lstlisting}[frame = t]
println 'f[x] = Sin[x]'.@\ttt@ >> 'f@\listingtildestr@(3)[x**2]'.@\ttt@
\end{lstlisting}
\begin{sout}
-Cos[x**2]
\end{sout}
In the case of multivariate functions, one should specify how many times to 
differentiate with respect to each slot (argument):
\begin{lstlisting}
def t = 'f@\listingtildestr@(2, 3, 0)[a**2, w, q]'.@\ttt@
\end{lstlisting}
represents \( \displaystyle \left. \frac{\partial^5}{\partial x^2 \partial y^3} f\left(x, y, q\right) \right|_{x = a^2 ,\, y =w}\). The above notation applies to derivatives of pure tensor fields. If one need to take derivative of some particular expression one should use the following syntax \inlinecode{D[var1, var2, ...][expression]}, which evaluates derivative of \inlinecode{expression} with respect to specified variables:
\begin{lstlisting}[frame = t]
println 'D[x, y, y][ y*f[x**2, y] ]'.@\ttt@
\end{lstlisting}
\begin{sout}
2*y*x*f@\listingtilde@(1, 2)[x**2, y] + 4*x*f@\listingtilde@(1,1)[x**2, y]
\end{sout}
More information on how to take derivatives can be found in Sec.~\ref{sec_Differentiate}.

In the case of indexed objects one should also specify additional indices of differentiation variables. For example, expression
\begin{lstlisting}
def t = 'F@\listingtildestr@(2)_{mn ab}^{cd}[f_ab]'.@\ttt@
\end{lstlisting}
represents \( \displaystyle \left. \frac{\delta}{\delta t^{ab}} \frac{\delta}{\delta t_{cd}} F_{mn}\left(t_{ab}\right) \right|_{t_{ab} = f_{ab}} \). As we see, the inverted indices of differentiation variables should be appended to the indices of pure tensor field. Since the relative ordering of derivatives is irrelevant, the indices of derivative have additional symmetries. From the previous example:
\begin{lstlisting}[frame=t]
println @
@t.@\property{indices}@.@\property{symmetries}@.@\property{permutationGroup}
\end{lstlisting}
\begin{sout}
Group( +[[2, 4], [3, 5]] )
\end{sout}
Indices of differentiation variables are considered to be appended sequentially starting from the first argument. So, for example, expression:
\begin{lstlisting}
def t = 'F@\listingtildestr@(1, 1)_{mnabcde}[f_abc, f_ab]'.@\ttt@
\end{lstlisting}
represents \( \displaystyle \frac{\delta}{\delta f^{abc}} \frac{\delta}{\delta f^{de}} F_{mn}\left(f_{abc}, f_{ab}\right) \) but not \( \displaystyle \frac{\delta}{\delta f^{cde}} \frac{\delta}{\delta f^{ab}} F_{mn}\left(f_{abc}, f_{ab}\right) \).

\section{Transformations}
\label{sec_BasicTransformations}
\subsection{Applying and manipulating transformations}
All transformations in Redberry are first-class objects (which means that they can be assigned to variables) and share a common way for their applying and manipulation. Consider the syntax used for applying transformation to mathematical expression:
\begin{lstlisting}[frame=t, numbers=left]
def tr = Expand
def t = '(A_k + B_k)*c'.@\ttt@
def r = tr >> t,
    l = t << tr
assert r == l
println r
\end{lstlisting}
\begin{sout}
c*A_k+c*B_k
\end{sout}
As it seen from the example, transformations are applied using left shift \inlinecode{<<} or right shift \inlinecode{>>} operator. It should be noted that both operators are left-associative so in order to apply several transformations subsequently, it is better to use a special \inlinecode{&} operator, which allows to join a set of transformations into a single one:
\begin{lstlisting}[frame = t, numbers=left,name=TrAnd]
def t = '(a + b)*c'.@\ttt@
def expandAndSubs = Expand & 'c = a + b'.@\ttt@
println expandAndSubs >> t
\end{lstlisting}
\begin{msout}
a*(a+b)+b*(a+b)
\end{msout}
\begin{lstlisting}[numbers=left,name=TrAnd]
def subsAndExpand= 'c = a + b'.@\ttt@ & Expand 
println subsAndExpand >> t
\end{lstlisting}
\begin{sout}
a**2+2*a*b+b**2
\end{sout}

Transformations (like tensors) are immutable in Redberry. Some transformations may take required or optional arguments using square brackets: 
\begin{lstlisting}[frame=t, numbers=left, name=TrBr]
def eliminateWhileExpand = Expand[EliminateMetrics]
println eliminateWhileExpand >> '(g_mn + d_m^a*g_na)*f^mn'.@\ttt@
\end{lstlisting}
\begin{msout}
2*f_m^m
\end{msout}
\begin{lstlisting}[numbers=left, name=TrBr]
def diff = Differentiate['x_m']
println diff >> 'x_m*x^m'.@\ttt@
\end{lstlisting}
\begin{sout}
2*x^m
\end{sout}
In this example, the \inlinecode{Expand} transformation takes an optional parameter, which is a transformation to be applied on each level of expand procedure. In contrast, the argument of \inlinecode{Differentiate} transformation is required. In both cases a new object will be created and assigned to a corresponding variable. The meaning of arguments is specific for each particular transformation and will be discussed in further sections.

\subsection[List of selected general-purpose transformations]{List of selected general-purpose transformations\footnote{the full list of common transformations can be found in \href{http://redberry.cc/documentation:guide:list_of_transformations}{online documentation}}}

\subsubsection{Substitutions}
\label{sec_Substitutions}
The most frequent transformation in all computations is a substitution. Here we shall discuss the usage aspects of substitutions, while the idea of the underlying algorithms can be found in Sec.~\ref{sec_InternalArchitecture}.

The very important feature of any tensors-oriented CAS is automatic relabelling of dummy indices in the case of dummy indices clash. Redberry takes care about it in all types of substitutions. Consider, for example, the following simple substitution:
\[ x = x_a{}^a \qquad\text{in}\qquad  (x f_a + y_a) (x f_b + z_b) \]
Here is a code to perform this substitution in Redberry:
\begin{lstlisting}[frame  = t]
def s = 'x = x_a^a'.@\ttt@
def t = '(x*f_a + y_a)*(x*f_b + z_b)'.@\ttt@
println s >> t
\end{lstlisting}
\begin{sout}
(x_d^d*f_a+y_a)*(x_c^c*f_b+z_b)
\end{sout}
As one can see, the appropriate relabelling was performed automatically. 

Redberry supports substitutions of tensor fields and automatically performs matching of the arguments of functions during substitution:
\begin{lstlisting}[frame = t]
def s = 'F_ij[x_m, y_m] = x_i*y_j'.@\ttt@
def t = 'T^ab*F_ab[p^a - q^a, p^a + q^a]'.@\ttt@
println s >> t
\end{lstlisting}
\begin{sout}
T^ab*(p_a-q_a)*(p_b+q_b)
\end{sout}
If tensor field depends on indexed argument, then some ambiguities arises when mapping the arguments. For example, the following substitution: \[  F_i (x_{mn}) \,=\, x_{ij} f^j \qquad \rightarrow \qquad F_k(x_i\,y_j) \] can be performed in two different ways: (a) matching $x_{ij} \to x_i y_j$ gives $x_k y_j f^j$, (b) matching $x_{ij} \to x_j y_i$ gives $y_k x_j f^j$. This is because the indices of product ($x_i\,y_j$) are not ordered by their nature. To explicitly specify the matching rule in such ambiguous situations, Redberry allows to enter the correspondence of indices in the field arguments:
\begin{lstlisting}[name = ll, frame = t, numbers = left]
def s = 'F_i[x_mn] = x_ik*f^k'.@\ttt@
def t = 'F_k[x_i*y_j]'.@\ttt@ //equivalent to F_k[x_i*y_j:_ij]
println s >> t
\end{lstlisting}
\begin{msout}
 x_k*y_a*f^a
 \end{msout}
\begin{lstlisting}[name = ll, numbers = left]
t = 'F_k[x_i*y_j:_ji]'.@\ttt@ 
println s >> t
\end{lstlisting}
\begin{sout}[frame = b]
 x_a*y_k*f^a
 \end{sout}
 
As well as simple tensor and tensor field substitutions, Redberry fully supports all other types of substitutions, taking into account both indices symmetries and indices contractions. Consider the following complicated example. Let's apply the following substitution
\[f_{m} + R_{bma}\,F^{ba} - R_{ljm}\,F^{lj} =  R_{bam}\,F^{ab}\]
to tensor
\[f_i + R_{ijk} F^{jk} + R_{ijk}\,F^{kj} - R_{kij}\,F^{jk},\]
where $R_{abc}$ is antisymmetric: \( R_{abc} \,=\, - R_{cba}\). It is easy to show that the result will be zero. In Redberry one can e.g. do
\begin{lstlisting}[frame = t]
@\addAntiSymmetry@ 'R_mnp', 2, 1, 0
def s = 'f_m + R_bma*F^ba - R_ljm*F^lj =  R_bam*F^ab'.@\ttt@
def t = 'f_i + R_ijk*F^jk + R_ijk*F^kj - R_kij*F^jk'.@\ttt@
println s >> t
\end{lstlisting}
\begin{sout}
0
\end{sout}
We see, that Redberry matched the l.h.s. of the substitution in tensor \inlinecode{t} and automatically reduced the resulting sum to the standard form, which, in turn, gave zero.

Redberry takes into account not only predefined symmetries of simple tensors, but also symmetries of any complicated expression, which arise from its structure. For example:
\begin{lstlisting}[frame = t]
def s = 'K_a * (A^ab - A^ba) = F^a*A_a^b'.@\ttt@
def t = 'K_p * (A^qp - A^pq) + F^b*A_b^q'.@\ttt@
println s >> t
\end{lstlisting}
\begin{sout}
0
\end{sout}
The result is zero since tensor $(A_{cb} - A_{bc})$ is antisymmetric. As well, Redberry takes care about symmetries of indexless objects:
\begin{lstlisting}[frame = t]
def c = 'Cos[a - b] = c'.@\ttt@,
    s = 'Sin[a - b] = s'.@\ttt@,
    t = 'x = Cos[b - a]**3 + Sin[b - a]**3'.@\ttt@
println (c | s) >> t
\end{lstlisting}
\begin{sout}
x = c**3 - s**3 
\end{sout}
As one can see, the built-in definitions of sine and cosine are set up to be odd and even respectively.

There is an important note on applying several substitutions at a time using the joining of transformations. Substitutions joined with \inlinecode{&} operator will be applied sequentially. However, sometimes it is necessary to apply several substitution rules "simultaneously". Consider the following example:
\begin{lstlisting}[frame=t]
def X2YandY2X = 'x=y'.@\ttt@ & 'y=x'.@\ttt@,
     X2YorY2X  = 'x=y'.@\ttt@ | 'y=x'.@\ttt@,
     tensor = 'x + 2*y'.@\ttt@    
println X2YandY2X >> tensor
\end{lstlisting}
\begin{msout}
3*x
\end{msout}
\begin{lstlisting}
println X2YorY2X >> tensor
\end{lstlisting}
\begin{sout}
y+2*x
\end{sout}
The first transformation (\inlinecode{X2YandY2X}) means just sequential applying of two provided substitutions, while the second ( \inlinecode{X2YorY2X}) performs both substitutions "simultaneously".


\subsubsection{Differentiate}
\label{sec_Differentiate}
This transformation allows to take derivatives with respect to indexed objects. Consider the examples:
\begin{lstlisting}[name = DifferentiateEx,%
numbers = left, frame = t]
def tensor = 'Sin[f_ab*f^ab]'.@\ttt@    
println Differentiate['f_mn'] >> tensor
\end{lstlisting}
\begin{msout}
2*Cos[f^{ab}*f_{ab}]*f^{mn}
\end{msout}
\begin{lstlisting}[name = DifferentiateEx, numbers = left]
//derivative of antisymmetric tensor
@\static{setAntiSymmetric}@('R_ab')
println Differentiate['R_ab'] >> 'R_mn'.@\ttt@
\end{lstlisting}
\begin{sout}
(1/2)*(d_{m}^{a}*d_{n}^{b} - d_{n}^{a}*d_{m}^{b})
\end{sout}
The \inlinecode{Differentiate} transformation takes care about dummies relabelling and symmetries of tensors. The following convention is adopted: 
\[ 
\frac{\delta\,T_{m_1\dots m_k}}{\delta\,T_{n_1\dots n_k}} 
=
\frac{1}{N} \, ( \delta^{n_1}_{m_1}\dots\delta^{n_k}_{m_k} \,+\, \mbox{\itshape
permutations}),
\]
where $N$ is a number of elements in the sum, and r.h.s of this expression have the same symmetries as l.h.s.

\subsubsection{EliminateMetrics} 
This transformation eliminates metric tensors and Kronecker deltas, which are contracted with other tensors:
\begin{lstlisting}[name = ContractEx,%
numbers = left, frame = t]
println EliminateMetrics >> 'g_nm*A^m*d^n_a'.@\ttt@
\end{lstlisting}
\begin{msout}
A_a
\end{msout}
\begin{lstlisting}[name = ContractEx, numbers = left]
tensor = ('g^mn*g^ab*g^gd*(p_g*g_ba + p_a*g_bg)*(p_m*g_dn + p_n*g_dm)').@\ttt@ 
//eliminate metrics in D dimensions                
println ((EliminateMetrics & 'd^a_a = D'.@\ttt@) >> tensor)
\end{lstlisting}
\begin{sout}
2*(1+D)*p^d*p_d
\end{sout}

\subsubsection{Expand}
There are several transformations which expand out products and integer powers of sums: \inlinecode{Expand}, \inlinecode{ExpandAll},\\
\inlinecode{ExpandNumerator} and \inlinecode{ExpandDenominator}. These well-known transformations needs no special explanation except the fact that they can take an additional transformations as an arguments to be applied on the each level of expand procedure. Consider the following example:
\begin{lstlisting}[name = ExpandEx, 
numbers = left,
frame = t]
def tensor = '(g_af*g_bc+g_bf*g_ac+g_cf*g_ba)*(T_d*T_e+g_de)*(g^db*g^ae + g^de*g^ab)'.@\ttt@
//eliminates metrics in four dimensions
def eliminate = EliminateMetrics & 'd^a_a = 4'.@\ttt@
//expand and then eliminate
def r1 = (Expand & eliminate) >> tensor
println r1
\end{lstlisting}
\begin{msout}
2*T_{c}*T_{f}+30*g_{fc}+7*g_{fc}*T_{e}*T^{e}
\end{msout}
\begin{lstlisting}[name = ExpandEx, 
numbers = left,
frame = b]
//eliminate while expand
def r2 = Expand[eliminate] >> tensor
assert r1 == r2
\end{lstlisting}
As we see, lines 6 and 9 produce the same result, but the latter spends less time in the calculation because it applies additional simplifications not only to the final result, but to all intermediate tensors, drastically reducing their complexity.

\subsubsection{Factor}
This transformation factors scalar polynomials over the integers in expressions\footnote{The code of polynomials factorization is taken from free and open source  \href{http://krum.rz.uni-mannheim.de/jas/}{Java Algebra System} \cite{Kredel2008185, Kredel08-0} designed by Heinz Kredel.}. Consider the examples:
\begin{lstlisting}[frame=t,numbers=left,name=FactorEx]
//multivariate polynomial
def t = '2*x**3*y - 2*a**2*x*y - 3*a**2*x**2 + 3*a**4'.@\ttt@
println Factor >> t
\end{lstlisting}
\begin{msout}
(x+a)*(x-a)*(-3*a**2+2*y*x)
\end{msout}

\begin{lstlisting}[numbers=left,name=FactorEx]
//tensorial expression with symbolic polynomial parts
t = Expand >> '(a+b)**4*F_mn + (x**6-y**6)*R_mn'.@\ttt@
println Factor >> t
\end{lstlisting}
\begin{msout}
(a+b)**4*F_mn+(x+y)*(x-y)*(x*y+x**2+y**2)*(-x*y+x**2+y**2)*R_mn
\end{msout}

\begin{lstlisting}[numbers=left,name=FactorEx]
//tensorial expression with scalar combinations
t = '(a+b)**2*f_m*f^m + (a**2 - b**2)*f_a*f^a*f_b*f^b'.@\ttt@
println Factor >> t
\end{lstlisting}
\begin{msout}
((a-b)*f_a*f^a+a+b)*(a+b)*f_m*f^m
\end{msout}

\begin{lstlisting}[numbers=left,name=FactorEx]
//do not factor "tensorial" scalars
println Factor[[@\mapitem{FactorScalars}@: false]] >> t
\end{lstlisting}
\begin{sout}
(a+b)**2*f_m*f^m-(b+a)*(b-a)*f_a*f^a*f_b*f^b
\end{sout}   
\inlinecode{Factor} traverses the expression from head to children and factors all scalar polynomials (both univariate and multivariate) and rational functions. So, in the case of scalar expression it applies only to the top algebraic level. In the case of rational expression, \inlinecode{Factor} first calls \inlinecode{Together} (Sec.~\ref{sec_Together}), then factors numerator and denominator.

As it seen from the example, \inlinecode{Factor} tries to factor all scalar combinations of tensors (like \inlinecode{f_m*f^m} in line 8). This can be changed by passing an additional option \inlinecode{FactorScalars:false} as it done in line 11.

It should be noted, that in the current version of Redberry, performance of \inlinecode{Factor} transformation in case of multivariate polynomials may be low in some cases and some huge expressions may not be factored completely. Also \inlinecode{Factor} applies only to expressions, which does not contain trigonometric functions, tensor fields or non integer powers of variables. These issues will be addressed in the upcoming releases of Redberry.

\subsubsection{Together}
\label{sec_Together}
There are two transformations, which put terms in a sum over a common denominator: \inlinecode{Together} and \inlinecode{TogetherFactor}. The last one also cancels scalar factors in the result. Consider the examples:
\begin{lstlisting}[frame = t, numbers = left, name=TogetherEx]
def t = 'x**2/(x**2 - 1) + x/(x**2 - 1)'.@\ttt@
//to common denominator and cancel factors
println TogetherFactor >> t
\end{lstlisting}
\begin{msout}
(1+x)**(-1)*(-1+x)**(-1)*(x+x**2)
\end{msout}

\begin{lstlisting}[numbers = left, name=TogetherEx]
//together with tensorial expressions
t = 'f_m/a + k_m/(f_m*f^m)'.@\ttt@
println Together >> t
\end{lstlisting}
\begin{sout}
a**(-1)*(f_{a}*f^{a})**(-1)*(f_{b}*f^{b}*f_{m}+a*k_{m})
\end{sout}
As it seen from the last example, \inlinecode{Together} effectively relabel
dummy indices when it is necessary.

\subsection[List of selected physical transformations]{List of selected physical transformations} 
\subsubsection{DiracTrace}
This transformation calculates the Dirac trace of expression in four dimensions\footnote{this code requires \inlinecode{cc.redberry.core.indices.IndexType.*} to be added to static imports}:
\begin{lstlisting}[frame=t, numbers=left, name=DiracTraceEx1]
//set up matrices
@\static{defineMatrices}@ 'G_a', 'G5', Matrix1.@\property{matrix}@ 
//setting up symmetries of Levi-Civita
@\static{setAntiSymmetric}@ 'e_abcd'
//dirac trace transformation
def dTrace = DiracTrace[[@\mapitem{Gamma}@: 'G_a',@
@ @\mapitem{Gamma5}@: 'G5', @\mapitem{LeviCivita}@: 'e_abcd']]
println dTrace >> 'Tr[G_a*G_b*G_c*G_d*G5]'.@\ttt@
\end{lstlisting}
\begin{msout}
-4*I*e_{abcd}
\end{msout}

\begin{lstlisting}[numbers=left, name=DiracTraceEx1]
println dTrace >> 'Tr[(p_a*G^a + m)*G_m*G5*(q_a*G^a-m)*G_n]'.@\ttt@
\end{lstlisting}
\begin{sout}
-4*I*p_{b}*q_{a}*e^{a}_{n}^{b}_{m}
\end{sout}
The first line tells Redberry to consider tensor \inlinecode{G_a} and \inlinecode{G5} as matrices with additional one upper and one lower index of type \inlinecode{Matrix1} (Latin lower letters with strokes), i. e. \inlinecode{G_a^i'_j'} and \inlinecode{G5^i'_j'} respectively. Then Redberry will use matrix multiplication rules for the products of such tensors\footnote{for further details on matrix objects in Redberry see \href{http://redberry.cc/documentation:guide:setting_up_matrix_objects}{online documentation}}. Line 6 defines \inlinecode{DiracTrace} transformation with specified notation for gamma matrices and Levi-Civita tensor.

\subsubsection{UnitaryTrace}
This transformation calculates traces of unitary matrices:
\begin{lstlisting}[frame=t, numbers=left, name=UnitaryTraceEx]
//set up matrices
@\static{defineMatrices}@ 'T_A', Matrix2.@\property{matrix}@
//structure constants are antisymmetric
@\static{setAntiSymmetric}@ 'f_ABC'
//d-constants are symmetric
@\static{setSymmetric}@ 'd_ABC'
//unitary trace transformation
def uTrace = UnitaryTrace[[@\mapitem{Matrix}@: 'T_A', @\mapitem{f}@:@
@'f_ABC', @\mapitem{d}@: 'd_ABC', @\mapitem{N}@: 'N']]
println uTrace >> 'Tr[T_A*T_B*T_C]'.@\ttt@
\end{lstlisting}
\begin{msout}
(1/4*I)*f_{CAB}+(1/4)*d_{CAB}
\end{msout}

\begin{lstlisting}[numbers=left, name=UnitaryTraceEx]
println uTrace >> 'Tr[T_A*T_B*T_C*T^A]'.@\ttt@
\end{lstlisting}
\begin{sout}
(-(1/4)*N**(-1)+(1/4)*N)*g_{BC}
\end{sout}
As one can see, it is necessary to specify the notation used for structure and
$d$- constants (\inlinecode{f_ABCD} and \inlinecode{d_ABCD}) and dimension
of unitary group (\inlinecode{N}).

\subsubsection{LeviCivitaSimplify}
This transformation simplifies combinations of Levi-Civita tensors:
\begin{lstlisting}[frame=t, numbers=left, name=LeviCivitaSimplifyEx]
//three dimensions
@\static{setAntiSymmetric}@ 'e_abc'
def t = 'e_abc*e^abd'.@\ttt@
//simplify in Euclidean space in three dimensions
println LeviCivitaSimplify.@\property{euclidean}@['e_abc'] >> t
\end{lstlisting}
\begin{msout}
2*d^d_c
\end{msout}

\begin{lstlisting}[numbers=left, name=LeviCivitaSimplifyEx]
//four dimensions
@\static{setAntiSymmetric}@ 'e_abcd'
t = '4*I*e^h_d^fb*e_abch*e_e^d_gf'.@\ttt@
//simplify in Euclidean space in four dimensions
println LeviCivitaSimplify.@\property{euclidean}@['e_abcd'] >> t
\end{lstlisting}
\begin{msout}
16*I*e_{eagc}
\end{msout}

\begin{lstlisting}[numbers=left, name=LeviCivitaSimplifyEx]
t = '4*I*e^h_d^fb*e_abch*e_e^d_gf'.@\ttt@
//simplify in Minkowski space in four dimensions
println LeviCivitaSimplify.@\property{minkowski}@['e_abcd'] >> t
\end{lstlisting}
\begin{sout}
-16*I*e_{eagc}
\end{sout}
As one can see, it is necessary to specify whether the space is considered to be Euclidean or Minkowski and the notation for Levi-Civita tensor. The difference between Euclidean and Minkowski Levi-Civita tensors appears for even number of space-time dimensions (compare lines 10 and 13). Additionally, the transformation automatically substitute the dimension of space, which is considered to be equal to the number of Levi-Civita indices.

\section{Selected physical applications}
\label{sec_PhysExamples}
\subsection{Feynman diagrams}
\label{sec_ComptonExpampleQED}
One of the practical applications of Redberry is calculation of Feynman diagrams in Quantum Field Theory. Redberry provides several common physical transformations such as traces of Dirac gammas and SU(N) matrices, simplification of Levi-Civita combinations etc. and powerful tools for simple inputting of noncommutative matrix expressions. In this section we shall demonstrate these features by calculation of well-known differential cross section of the Compton scattering in quantum electrodynamics (i.e. spinor electrodynamics, in contrast to the scalar electrodynamics, which was illustrated in Sec.~\ref{sec_IntroductoryExample}).

First of all it is necessary to note, that Redberry does not support noncommutative products of indexless variables. However, many noncommutative objects, which occur in physical calculations (e.g. spinors or Gamma matrices) have matrix origin, which means, that they have additional matrix indices, which are usually omitted for convenience. For example, when we write product of Dirac bispinors and gammas, like e.g.
\begin{equation}
\label{eq:Matrix1}
T_{\mu\nu} = \bar u \, \gamma_{\mu} \gamma_{\nu}
\end{equation}
we actually mean, that the above quantities have additional matrix indices, which 
can be written explicitly:
\begin{equation}
\label{eq:Matrix2}
T_{\mu\nu\,b'}  = \bar u_{a'} \, \gamma_{\mu}{}^{a'}{}_{c'}
\gamma_{\nu}{}^{c'}{}_{b'}.
\end{equation}
In this expression \(a',b',c'\) are special matrix indices. From such a point of view these "noncommutative" products can be represented as ordinary products of indexed objects. Such matrix indices should have a nonmetric type, which implies that it is impossible to perform raising or lowering or define a symmetry which mixes indices with different states. Redberry provides an internal facilities allowing to input matrix expressions in a convenient form like (\ref{eq:Matrix1}). Matrix indices will automatically be inserted at parsing and subsequent processing could be made as with usual indexed expressions. Detailed description of this feature can be found in \href{http://redberry.cc/documentation:guide:setting_up_matrix_objects}{online documentation}.

So, let us turn to the physical aspects of the problem. The following lines give squared matrix element of the Compton scattering in quantum electrodynamics\footnote{\inlinecode{cc.redberry.core.indices.IndexType.*} should be added to static imports}:
\begin{lstlisting}[frame = t, name = ComptonQED, numbers = left ]
@\static{defineMatrices}@ 'G_a', 'V_i', 'D[x_m]', Matrix1.@\property{matrix}@,
            'vu[p_a]', Matrix1.@\property{vector}@,
            'cu[p_a]', Matrix1.@\property{covector}@
//photon-electron vertex
def V = 'V_m = -I*e*G_m'.@\ttt@,
    //electron propagator
    D = 'D[p_m] = -I*(m + p_m*G^m)/(m**2 - p_m*p^m)'.@\ttt@,
    //diagram a)
    Ma= 'cu[p2_m]*V_m*e^m[k2_m]*D[k1_m+p1_m]*V_n*e^n[k1_m]*vu[p1_m]'.@\ttt@,
    //diagram b)
    Mb= 'cu[p2_m]*V_m*e^m[k1_m]*D[p1_m-k2_m]*V_n*e^n[k2_m]*vu[p1_m]'.@\ttt@,
    //matrix element
    M = (V & D) >> (Ma + Mb)
//Mandelstam variables
def mandelstam = @\static{setMandelstam}@(['p1_m': 'm', 'k1_m': '0', 'p2_m': 'm', @
@'k2_m': '0'])
M = (ExpandAll & mandelstam) >> M
//complex conjugation
def MC = M
MC = 'vu[p1_m]*cu[p2_m] = vu[p2_m]*cu[p1_m]'.@\ttt@ >> MC
MC = (Conjugate & Reverse[Matrix1]) >> MC
//squared matrix element
def M2 = ExpandAll >> (M * MC / 4)
//photon polarizations
M2 = 'e_m[k1_a]*e_n[k1_a] = -g_mn'.@\ttt@ >> M2
M2 = 'e_m[k2_a]*e_n[k2_a] = -g_mn'.@\ttt@ >> M2
//electron polarizations
M2 = 'vu[p2_m]*cu[p2_m] = m + p2^m*G_m'.@\ttt@ >> M2 
M2 = 'vu[p1_m]*cu[p1_m] = m + p1^m*G_m'.@\ttt@ >> M2
//applying trace of gamma matrices
M2 = DiracTrace['G_a'] >> M2
//final simplifications
M2 = (ExpandAndEliminate & 'd^m_m = 4'.@\ttt@ & mandelstam) >> M2
M2 = ('u = 2*m**2 - s - t'.@\ttt@ & Factor) >> M2
println M2
\end{lstlisting}
\begin{sout}
2*(-m**2+s+t)**(-2)*e**4*(-8*s**2*m**2*t+4*s**3*t+2*s**4+t**3*s+2*m**8+4*m**4*s@
@*t-m**2*t**3-2*m**2*t**2*s+3*m**4*t**2-8*s**3*m**2+12*s**2*m**4+3*s**2*t**2-8*@
@m**6*s)*(m**2-s)**(-2)
\end{sout}

The above code reproduces the standard steps of Feynman diagrams calculation and prints squared matrix element of the Compton scattering averaged over polarizations of initial particles and summed over polarizations of final particles:
\begin{multline*}
\frac{1}{4} \sum_{\mbox{spins}} \,|\mathcal M|^2 \,=\, \frac{2\,e^4
}{(m^{2}-s)^2 (-m^{2}+s+t)^2}
\,  \times \, \left( -\,8 s^{2} m^{2} t+4 s^{3} t+2 s^{4}+t^{3} s+2 m^{8}+4 m^{4} s t-m^{2} t^{3} 
\right. \\ \left.
-\,2 m^{2} t^{2} s+3 m^{4} t^{2}-8 s^{3} m^{2}+12s^{2} m^{4}+3 s^{2} t^{2}-8 m^{6} s \right)
\end{multline*}

Let's consider the code in more detail. Firstly, in lines 1-3 we tell Redberry to consider some input objects as matrices, vectors or covectors. These objects are: gamma matrices (\inlinecode{G_a}), photon-electron vertex (\inlinecode{V_i}), electron propagator (\inlinecode{D[p_m]}), electron wave function \(u(p)\) (\inlinecode{vu[p_m]}) and its conjugation \(\bar u(p)\) (\inlinecode{cu[p_m]}). To make this, \inlinecode{defineMatrices(...)} function is invoked with lists of tensors followed by target object descriptors. Each descriptor specifies the type of matrix indices and a type of matrix (e.g. \inlinecode{Matrix1.matrix}, \inlinecode{Matrix1.covector}). Here only one matrix index type is used. As a result, for example, the combination like \( \bar u(p_1) \, \gamma_\mu \, u(p_2)\) will be treated as a scalar (with respect to matrix indices), while the combination like \(u(p_1) \, \bar  u(p_2)\) will be treated as a matrix. In expressions like $m + p_\mu \gamma^\mu$ first term will automatically be multiplied by the identity matrix (Kronecker delta).

The next step is to build the matrix element. As it well known, two Feynman diagrams corresponds to the Compton scattering in the leading order. This is done in lines 9 - 13, where 
\inlinecode{e_n[k_m]} denotes photon polarization. In line 15 we define Mandelstam variables. Function
\begin{lstlisting}
@\static{setMandelstam}@(['p1':'m1', 'p2':'m2', 'p3':'m3', 'p4':'m4'])
\end{lstlisting}
takes a map 'momentum --- mass of particle' as an argument and returns a collection of substitutions, followed from the usual Mandelstam definitions for  incoming momentums \inlinecode{p1} and \inlinecode{p2} and outcoming momentums \inlinecode{p3} and \inlinecode{p4}.

The next step (lines 18 - 20) is to make up a complex conjugation of the matrix element. As it is known from physics, this can be done using the following relation:
\[
\left( \bar u(p_2) \, \hat M \, u(p_1) \right)^* =  \bar u(p_1) \, \hat{ \bar M
}\, u(p_2) ,
\,\,\mbox{where}\quad
\hat M = \sum_i c_i \, \hat a_1 \, \hat a_2 \, \dots \, \hat a_i 
\,\,\mbox{,}\quad
\hat {\bar M }= \sum_i c_i^* \, \hat a_i \, \hat a_{i-1} \, \dots \, \hat a_1
\]
and usual notation \(\hat a = \gamma_{\mu} a^{\mu}\) is used. This transformation is equivalent to two substitutions:
\[
u(p_1) \, \bar u(p_2) \to u(p_2) \, \bar u(p_1)\quad\mbox{and}\quad
\hat M \to \hat {\bar M },
\]
which are applied in line 19 and 20 respectively. The meaning of \inlinecode{Reverse[Matrix1]} transformation is clear from its name: it simply overwrites the matrix product in the reverse order.

At this point we can define squared matrix element in line 22 and perform summation over photons  (lines 24 - 25) and electrons (lines 27 - 28) polarizations. On the last step all products of gamma matrices are automatically converted to combinations of their traces. So the next step is to apply \inlinecode{DiracTrace} transformation. After this, it is necessary to eliminate metric tensors and apply Mandelstam substitutions (line 32). This step will produce a symbolic expression, which then can be finally simplified using \inlinecode{Factor} transformation. The whole calculation takes less than $200$ milliseconds on the warmed Java Virtual Machine (on Intel Core i5 @ 2.27 Ghz).

\subsection{One-loop counterterms of arbitrary Lagrangians}
\label{sec_OneLoopExa}
Another effective application of Redberry is calculation of the divergent part of the one-loop effective action of arbitrary Lagrangians. The theoretical formalism based on the extended t\rq Hooft and Veltman method of background calculations was developed in \cite{Pronin:1996rv} and successfully applied for a number of theories using \textsc{Reduce} computer algebra system \cite{Pronin:1995}. This algorithm calculates one-loop counterterms for an arbitrary theory and background in four dimensions in curved space-time in the dimensional regularization. Redberry implements this algorithm for second and fourth order differential operators and provides a simple and convenient user interface.

Let's start with a minimal physical background. It is well-known, that one-loop effective action for a general field theory with a given action $S[\phi]$ can be expressed in terms of derivative of action with respect to the fields:
\[
\Gamma^{(1)} = \frac{i\hbar}{2} \, \mbox{Tr}\left( \, \ln \frac{\delta^2 
S}{\delta \phi^{i}
\delta \phi_j}\right),
\]
where $\phi_{i}$ denotes a fields and Latin letters denotes the whole set of its indices\footnote{Here the actual type of these indices is not important (e.g. this set can contain both space-time and SU(N) indices), so Latin letters used only for convenience.}. So, the main quantity, which determines the effective action is a differential operator
\begin{equation}
\label{eq:DiffOperator}
D_{i}{}^{j} \,=\, \frac{\delta^2 S}{\delta \phi^{i} \delta \phi_j}.
\end{equation}
In the most general case, this operator has the following form:
\begin{multline}
\label{eq:OperatorGenForm}
D_{i}{}^{j}\,= \, K^{\mu_1\mu_2\dots\mu_L}{}_i{}^j\,
\nabla_{\mu_1}\nabla_{\mu_2}\dots\nabla_{\mu_L}
+\, S^{\mu_1\mu_2\dots\mu_{L-1}}{}_i{}^j\,
\nabla_{\mu_1}\nabla_{\mu_2}\dots\nabla_{\mu_{L-1}} \,+
\\
+\, W^{\mu_1\mu_2\dots\mu_{L-2}}{}_i{}^j\,
\nabla_{\mu_1}\nabla_{\mu_2}\dots\nabla_{\mu_{L-2}} 
+\, N^{\mu_1\mu_2\dots\mu_{L-3}}{}_i{}^j\,
\nabla_{\mu_1}\nabla_{\mu_2}\dots\nabla_{\mu_{L-3}}
+\, M^{\mu_1\mu_2\dots\mu_{L-4}}{}_i{}^j\,
\nabla_{\mu_1}\nabla_{\mu_2}\dots\nabla_{\mu_{L-4}} \,+\,\dots,
\end{multline}
where $\nabla_\mu$ is a covariant derivative with respect to space-time and
internal indices:
\begin{eqnarray*}
&&\nabla_{\alpha} T^{\mu}{}_i{}^j = \partial_{\alpha}T^{\mu}{}_i{}^j +
\Gamma^\mu_{\alpha\gamma} \, T^\gamma{}_i{}^j + \omega_{\alpha i}{}^k
T^\mu{}_k{}^j - \omega_{\alpha k}{}^j T^\mu{}_i{}^k\\
&&\nabla_{\mu} \Phi_i = \partial_\mu \Phi_i + \omega_{\mu i}{}^j \Phi_j,
\end{eqnarray*}
where $\Gamma^\mu_{\alpha\gamma}$ is a Christoffel symbol and $\omega_{\mu i}{}^j$ is a connection on the principal bundle. Commuting covariant derivatives it is always possible to make tensors K, S, W, N, M symmetric in the Greek indexes and we shall assume this condition in the further reading.

It is also required to introduce the following quantities:
\begin{equation}
\label{eq:DefKn}
(Kn)^{i}_{j} =
K^{\mu_{1}\mu_{2}\dots\mu_{L}}{}^i{}_j\,\,n_{\mu_1}n_{\mu_2}\dots n_{\mu_L},\\
\qquad
(Kn)^{-1}{\,}_{i}{}^{j} \, (Kn)_j{}^k \,=\, \delta_i{}^k,
\end{equation}
where \(n_\mu\) is a unit vector. The second equation defines tensor \((Kn)^{-1}\) inverse to tensor \(Kn\), which is an input tensor for the algorithm\footnote{Redberry also provides facilities to find out the inverse tensor from the given equation: see examples on \href{http://redberry.cc/documentation:ref:reduce}{Redberry website}.}. The other required input is a curvature tensor with respect to the principal bundle:
\begin{equation}
\label{eq:FCurvature}
[\nabla_\mu, \nabla_\nu] \Phi_i = F_{\mu\nu}{}_i{}^j \Phi_j
\end{equation}

Given a set of tensors $K$, $W$, $M$, $(Kn)^{-1}$ and $F$ as input data,  Redberry allows to calculate counterterms for the general second and fourth order operators of the following form
\begin{eqnarray}
\label{eq:DiffOperator2}
&&D^{(2)}{}_{i}{}^j = K^{\mu\nu}{}_i{}^j\,\nabla_{\mu}\nabla_{\nu} + W_{i}{}^j\\
\label{eq:DiffOperator4}
&&D^{(4)}{}_{i}{}^j =
K^{\mu\nu\alpha\beta}{}_i{}^j\,\nabla_{\mu}\nabla_{\nu}\nabla_{\alpha}\nabla_{
\beta} + W^{\mu\nu}{}_i{}^j\,\nabla_{\mu}\nabla_{\nu} + M_{i}{}^j
\end{eqnarray}
Let us illustrate the usage by the particular examples.

\subsubsection{Vector field operator}
Let us consider one-loop counterterms of the the vector field operator, which appears in the theory of the massive vector field:
\[
D_\alpha{}^\beta = 
\delta_\alpha{}^\beta \Box  - \lambda \, \nabla_\alpha \nabla^\beta +
P_\alpha{}^\beta,
\]
where \(\Box =g^{\mu\nu}\nabla_\mu\nabla_\nu\) and \(\lambda = 1 + 1/\xi\). This is a second order operator, and in order to rewrite it in the form (\ref{eq:DiffOperator2}), it is necessary to symmetrize the second term by commutation of the covariant derivatives:
\[
D_\alpha{}^\beta = \left( g^{\mu\nu} \delta_\alpha^\beta - \frac{\lambda}{2}
\left( g^{\mu\beta}\delta_\alpha^\nu + g^{\nu\beta}\delta_\alpha^\mu \right)
\right) \nabla_\mu\nabla_\nu + P_\alpha{}^\beta +
\frac{\lambda}{2}R_\alpha{}^\beta,
\]
where $R_{\alpha\beta}$ is the Ricci tensor.

Using equations (\ref{eq:DefKn}) it can be easily found that 
\begin{equation*}
(Kn)_{\alpha}{}^\beta = \delta_{\alpha}^\beta - \lambda \, n_\alpha n^\beta,
\qquad
(Kn)^{-1}{}_{\alpha}{}^\beta = \delta_{\alpha}^\beta + \frac{\lambda}{1-
\lambda} n_\alpha n^\beta.
\end{equation*}

Hereby, at this point we have whole set of input tensors required by the algorithm:
\begin{eqnarray*}
&&K^{\mu\nu}{}_\alpha{}^\beta \,=\, g^{\mu\nu} \delta_\alpha^\beta -
\frac{\lambda}{2}\left( g^{\mu\beta}\delta_\alpha^\nu +
g^{\nu\beta}\delta_\alpha^\mu \right)\,,
\quad S^\mu{}_\alpha{}^\beta \,=\, 0\,,
\quad W_\alpha{}^\beta \,=\, P_\alpha{}^\beta + 
\frac{\lambda}{2}R_\alpha{}^\beta\,,
\\&& (Kn)^{-1}{}_{\alpha}{}^\beta \,=\, \delta_{\alpha}^\beta + 
\frac{\lambda}{1-
\lambda} n_\alpha n^\beta\,,
\quad F_{\mu\nu}{}_\alpha{}^\beta \,=\, R_{\mu\nu}{}_\alpha{}^\beta\,.
\end{eqnarray*}
In the further calculations we shall use the definition $\lambda = \gamma/(1+\gamma)$ for convenience (so $\gamma = \lambda/(1-\lambda)$).

The following code calculates one-loop counterterms of the vector field theory in curved space-time (here \inlinecode{g} used for $\gamma$):
\begin{lstlisting}[frame = t, numbers = left]
@\static{addSymmetries}@ 'R_abcd', -[1, 0, 2, 3].@\ppp@, [2, 3, 0, 1].@\ppp@
@\static{setSymmetric}@ 'R_ab', 'P_ab'
def iK = 'iK_a^b = d_a^b + g*n_a*n^b'.@\ttt@
def K = 'K^mn_a^b = g^mn*d_a^b - g/(2*(1+g))*(g^mb*d_a^n + g^nb*d_a^m)'.@\ttt@
def S = 'S^r^m_n = 0'.@\ttt@
def W = 'W^a_b = P^a_b+g/(2*(1+g))*R^a_b'.@\ttt@
def F = 'F_mnab = R_mnab'.@\ttt@
//calculates one-loop counterterms of the second order operator
def div = oneloopdiv2(iK, K, S, W, F)
def counterterms = 'P^a_a = P'.@\ttt@ >> div.counterterms[1]
counterterms = Collect['R', 'P', Factor[[@\mapitem{FactorScalars}@: false]]] >> @
@counterterms
println counterterms
\end{lstlisting}
\begin{sout}
counterterms = @
@(1/240)*R**2*(5*g**2+20*g+28)+(1/24)*(6*g+12+g**2)*P^{b}_{a}*P^{a}_{b}@
@+(1/120)*(5*g**2+10*g-32)*R_{fp}*R^{fp}@
@+(1/48)*P**2*g**2+(1/24)*(2*g+4+g**2)*P*R+(1/12)*g*(4+g)*R_{a_{5}}^{b}*P^{a_{5}}@
@_{b}
\end{sout}
In order to obtain one-loop counterterms in the dimensional regularization, one should multiply the result produced by Redberry by $1/16\pi(d-4)$ and integrate it over the space-time volume:
\begin{multline*}
\Gamma^{(1)}_{\infty} = \frac{1}{16\pi(d-4)} \int d^4 x  \sqrt{-g} \left(  
\frac{1}{120}(-32+5 \gamma^2+10 \gamma) R_{\epsilon\mu} R^{\epsilon\mu}
+\frac{1}{48}\gamma^2 P^2 +\right.\\
+\frac{1}{240} R^2 (28+5 \gamma^2+20 \gamma)+\frac{1}{24} (\gamma^2+12+6 \gamma)
P_{\beta\alpha} P^{\alpha\beta} +\\
\left.+\frac{1}{12}\gamma (4+\gamma) R_{\nu\epsilon} P^{\nu\epsilon}
+\frac{1}{24} R (\gamma^2+4+2 \gamma) P\right)
\end{multline*}

The above code is clear enough, but some remarks are needed. First of all, the main method \inlinecode{oneloopdiv2(...)}, which calculates the counterterms of the second order operator, returns a special object, which holds some intermediate results (like e.g. $RR$, $RF$, $FF$ parts of effective action from \cite{Pronin:1996rv}). The whole result can be obtained by getting the value of \inlinecode{.counterterms} property. All input expressions must be in the same notation as in the original work \cite{Pronin:1996rv} except tensor $(Kn)^{-1}$, which should be denoted as \inlinecode{iK}. At this moment, the implementation requires that all indices of input tensors must be lower Latin indices. Also, it assumed that field indices are placed at the end, so, for example, the first two indices of tensor $K_{\mu\nu}{}^\alpha{}_\beta$ are contracted with covariant derivatives in (\ref{eq:OperatorGenForm}), while the last two indices corresponds to the indices of vector field. Redberry does not support nonzero tensor $S$ from (\ref{eq:OperatorGenForm}), however it should be specified explicitly, like it is done in line 11.

\section{Basic internal architecture}
\label{sec_InternalArchitecture}
\subsection{Mappings of indices}
\label{sec_IndexMapping}
Perhaps the most significant difference between tensor- and symbol-oriented computer algebra systems lies in the comparison of mathematical expressions. In the symbol-oriented CASs result of  atomic comparison problem\footnote{determination of whether two expressions are equal, i.e. operation that is the main building block of such complicated routines as pattern matching} is just a logical true or false, while in the case of tensor-oriented CAS it transforms into a complicated pattern matching problem, which produces a complicated object as a result. The comparison problem of tensorial expressions will be revealed in this section through several examples.

The most common question, which can be asked about two expressions, is whether they define the same tensor (to within a free indices relabelling). This question arises in such frequent routines like substitutions and reduction of similar terms. Consider the following expressions:
\[
F_{ab}G^{bc} \mappingarrow F_{iq}G^{qj} \,\,=\,\, \left\{
    \begin{array}{c}
    a \to i\\
    c \to j
    \end{array}
  \right\}
\]
These two expressions have the same tensorial structure. The above notation means that if rename $a$ to $i$ and $c$ to $j$ in the left expression one will get exactly the same tensor as defined by the right expression. So, the result of such comparison is not just true or false, but a {\itshape mapping of free indices} of one expression onto free indices of another expression.


One of the major quirks of the problem lies in the fact that free and dummy indices hold completely different places. Mappings of {\itshape free} indices are global for expression, while {\itshape dummy} indices have their scopes. If some free index is present in several places, then its mapping will be the same everywhere. On the other hand, explicit names of dummies are not important (consider indices $b$ and $q$ in the first example). Besides that, dummy indices brings additional structure into expressions, which should be taken into account when finding mappings. To illustrate this features, consider the following examples:
\[
F_{{\color{blue}a}b}G^b{}_{\color{red}c}+M_{{\color{blue}a}d}N^d{}_{\color{red}c
}
\mappingarrow
F_{{\color{blue}i}q}G^q{}_{\color{red}j}+M_{{\color{blue}i}q}N^q{}_{\color{red}j
} 
\,\,=\,\, \left\{
    \begin{array}{c}
    a \to i\\
    c \to j
    \end{array}
  \right\}
\quad \mbox{but} \quad
F_{{\color{blue}a}b}G^b{}_{\color{red}c}+M_{{\color{blue}a}d}N^d{}_{\color{red}c
}
\mappingarrow
F_{{\color{blue}i}q}G^q{}_{\color{red}j}+M_{q{\color{blue}i}}N^q{}_{\color{red}j} 
\,\,=\,\, \varnothing.
\]
In the first expression $a$ and $c$ should be renamed into $i$ and $j$ respectively in both summands in order to transform l.h.s. into the r.h.s. But there is no mapping in the second example because structure of contractions of the second summand in the r.h.s. differs from that in the l.h.s. (if no symmetries defined for tensor $M_{ab}$).

\subsubsection{Multiple mappings and symmetries of tensors}
In general, several mappings of indices can exist for a pair of tensors. Consider the following primitive example. Suppose that tensor $R_{ab}$ is antisymmetric, then:
\[
R_{ab}A_c+R_{bc}A_a \mappingarrow R_{ij}A_k+R_{jk}A_i
\]
gives two mappings
\[
  \mathcal M_1=+\left\{
    \begin{array}{c}
    a \to i\\
    b \to j\\
    c \to k
    \end{array}
  \right\}\quad\mbox{and}\quad
  \mathcal M_2=-\left\{
    \begin{array}{c}
    a \to k\\
    b \to j\\
    c \to i
    \end{array}
  \right\}.
\]
Second mapping $\mathcal M_2$ has negative sign, which means that in order to obtain the r.h.s., one needs to apply mapping to the l.h.s. and negate the result. Sign property of  mappings and processing both symmetries and antisymmetries in a common way makes these entities fully consistent with each other.

It is clear that mapping of tensor onto itself gives permutational symmetries of its indices. So, in the case of the above primitive example, one can find that 
\[
R_{ab}A_c+R_{bc}A_a \mappingarrow R_{ab}A_c+R_{bc}A_a \,\,=\,\,
+\left\{
    \begin{array}{c}
    a \to a\\
    b \to b\\
    c \to c
    \end{array}
  \right\}\,\,\mbox{and}\,\,
 -\left\{
    \begin{array}{c}
    a \to c\\
    b \to b\\
    c \to a
    \end{array}
  \right\}.
\]
The last mapping represents a nontrivial antisymmetry of tensor. 

\subsubsection{Mappings in Redberry}
The entire architecture of Redberry rely on the ideas described in previous subsections. In Redberry one can construct mappings using the following syntax:
\begin{lstlisting}[name = IndexMappings, frame = t, numbers = left]
@\static{setAntiSymmetric}@ 'R_ab'
def from = 'R_ab*A_c + R_bc*A_a'.@\ttt@, to = 'R_ij*A^k + R_j^k*A_i'.@\ttt@
def mappings = from % to
println mappings.@\property{first}@ //takes only first mapping
\end{lstlisting}
\begin{msout}
+{_a->_i, _b->_j, _c->^k}
\end{msout}
\begin{lstlisting}[name = IndexMappings, numbers = left]
//print all mappings
mappings.each { mapping ->
   assert (mapping >> from) == to //apply mapping
   println mapping  
}
\end{lstlisting}
\begin{sout}
+{_a->_i, _b->_j, _c->^k}
-{_a->^k, _b->_j, _c->_i}
\end{sout}
Here, the construction \inlinecode{(from \% to)} gives a special object which allows to take just one mapping (this is very fast operation) using \inlinecode{.first} property (line 4) or to iterate over all possible mappings (line 6). Moreover, calculation of each subsequent mapping occurs only on the corresponding step of iteration (calculation on demand). In order to apply mapping rules to tensor one can use \inlinecode{>>} operator (as in line 7). This will automatically perform raising or lowering of indices if it is meant by mapping and resolve dummy clashes.

Redberry can build mappings for tensors of any complexity with any number of nested sums/products, symmetries and dummy indices:
\begin{lstlisting}[frame = t, numbers = left]
@\static{setAntiSymmetric}@ 'A_mn', 'F_mnab'
def from = '(A_m^n - A_m^p*A_p^n)*F_nk^i_j + A_mn*A^n_j*A^i_k'.@\ttt@,
    to = '-(A_d^a + A_p^a*A_d^p)*F^d_kq^i - A^a_b*A^b_q*A^i_k'.@\ttt@
(from % to).each { println it }
\end{lstlisting}
\begin{sout}
-{_i->_i, _j->_q, _k->_k, _m->^a}
+{_i->^k, _j->_q, _k->^i, _m->^a}
\end{sout}

\subsection{Graph isomorphism versus indices canonicalization}
\label{sec_versus}
One of the applications of finding mappings of indices is testing whether two expressions are equal (to within dummy indices relabelling and interchanging indices of simple tensors according to their symmetries). Probably, this operation is the most frequent operation that arises in any calculation (like reduction of similar terms). This problem can be solved by finding mapping of indices which maps free indices of one expression onto exactly same free indices of another.

As it was mentioned in the introduction, contractions between indices in product constitutes a graph. So, the problem of finding mappings between two expressions is equivalent to problem of testing whether two graphs are isomorphic. While no worst-case polynomial-time algorithms are known for the general Graph Isomorphism (GI) and closely related Graph Automorphism (GA) problems, there are several algorithms which solve these problems very effectively in all practical cases (see e.g. \cite{McKay1} and overview given in \cite{McKay2}). It is important to note, that the complexity of these algorithms (which is still exponential in the worst case) is given in the number of graph vertices, or equivalently in the number of product multipliers.

On the other hand, the existing systems (known to the authors) are based on the so-called {\itshape indices canonicalization approach} described in \cite{RodionovTaranov,Portugal}. This procedure brings indices of products into unique canonical order using the information about symmetries of its multipliers; when indices of tensors are brought into such order, it becomes trivial to test whether two expressions are equal. As it is shown in cited papers, this problem is equivalent to the problem of {\itshape double coset enumeration} which is known to be $\mathcal{NP}$-hard (see Sec.~4.6.8 in \cite{Holt}). Unfortunately, no really satisfactory algorithm for solving this problem has been found to date (see Sec.~4.6.8 in \cite{Holt}). Moreover, in contrast to GI problem, the complexity of algorithms for double coset enumeration is given in the number of all  indices (free and dummy) of tensor. It is also important, that indices canonicalization is not useful for testing whether two tensors are equal to within {\itshape free indices} relabelling. Additionally, indices canonicalization works only with products of simple tensors, so if e.g. product contains a sum one need to expand out brackets first.

Redberry utilises the first approach, based on the GI finding. The main algorithm, which finds mappings of tensor indices is actually searches for isomorphisms of corresponding graphs. Currently, this algorithm is far from the best known algorithms for GI problem. However, it is specifically suited for typical problems arising in physics and performs well on typical input  (i.e. when need to compare a huge number of small graphs; see Sec.~\ref{sec_comparison_performance}), but still has exponential complexity for some special cases. Nevertheless, there is a wide area for improvements and this is the main issue of further Redberry development.

Another important subject closely related to tensor comparison, is finding symmetries of complicated tensors. It is clear that from the stand point of graph-theoretical approach, this problem is equivalent to Graph Automorphism problem. This problem is also related to testing whether the expression is zero using only the information about its symmetries (like \(\displaystyle F_{ab}{}F^{bc}{}F_{c}{}^{a} = 0 \)  if $F_{ab}$ is antisymmetric). The canonicalization procedure used by other systems can naturally determine whether the expression is zero, while in Redberry one should apply special transformation \inlinecode{EliminateDueSymmetries}. The algorithm used in Redberry computes automorphism group $Aut(G)$ of corresponding graph $G$ and searches in this group for two equal permutations with different signs: if generators of group can produce two equal permutations with different signs (i.e. symmetry and antisymmetry), that means that corresponding expression is zero. Currently, Redberry uses a brute force algorithm for \inlinecode{EliminateDueSymmetries} and its performance is about of order of group, i.e. the number of all group elements $|Aut(G)|$, so it is impractical for large symmetric tensors. A vastly improved algorithm is under development and will appear soon in the next Redberry release.

Summarizing this section we point out that in our opinion, the approach based on graph algorithms provides a more flexible way to deal with tensors than canonicalization and has a broader perspective for further performance improvements. The good argument in favour of this point is a performance comparison of Redberry and other systems presented in Sec.~\ref{sec_comparison_performance}.

\subsection{Expression-tree traversal and modification}
As mentioned in Sec.~\ref{sec_architecture} each tensor in Redberry is a container of its child tensors, so any complicated expression becomes a hierarchical tree of such containers. Iteration over direct child elements of a tensor described in Sec.~\ref{sec_architecture}. Besides, there are a special tools for iteration and modification over a whole tree.

There is a core class in Redberry that performs traversal over any given expression. It basically generates a sequence of traversal events like: entering or leaving of subexpressions. The main feature of this class is its ability to in-place modify a tree while traversing. Redberry provides simple facade classes for tree traversal. There are two basic ways to perform a depth-first search on expression tree:
\begin{lstlisting}[frame = t, name = TreeIterator1, numbers = left]
def t = 'a + Sin[x + y]'.@\ttt@
t.parentAfterChild { a -> print a.toString() + ', ' }
\end{lstlisting}
\begin{msout}
a, y, x, y+x, Sin[y+x], a+Sin[y+x],
\end{msout}
\begin{lstlisting}[name = TreeIterator1, numbers = left]
t.parentBeforeChild { a -> print a.toString() + ', ' }
\end{lstlisting}
\begin{sout}
a+Sin[y+x], a, Sin[y+x], y+x, y, x,
\end{sout}
This example illustrates steps of parent-after-child and parent-before-child iteration modes.

Modification of expression tree can be performed in the same way. As an example, lets consider a naive implementation of a substitution transformation:
\begin{lstlisting}[frame=t, numbers=left,name = TreeIterator2]
def subs = { t, expr ->
  t.transformParentAfterChild {
     def mapping = expr[0] % it
     mapping.@\property{exists}@ ? mapping >> expr[1] : it
  }
}

def t = 'z_m*Cos[x_m*y^m - x_n*(z^n + t^n)] + t_m'.@\ttt@
println subs(t, 'z_a + t_a = y_a'.@\ttt@)
\end{lstlisting}
\begin{sout}
y_m
\end{sout}
The key aspect of parent-after-child modification, is that if child node is changed, then its parent node will be reduced to standard form (see Sec.~\ref{sec_StandardForm}) before it will be shown\footnote{The first modification will be performed on the sum \inlinecode{z^n + t^n}, which will be replaced with \inlinecode{y^n}. After this modification, the whole branch of the expression tree, which contains this term, will be permanently reduced to the standard form. So, when iteration will reach the argument of cosine, which will be \inlinecode{x_m*y^m - x_n*y^n}, it will be reduced to zero. On the next step the cosine receives zero as an argument and reduces to 1. On the last step of the iteration, it finally becomes possible to replace obtained \inlinecode{z_m + t_m} with \inlinecode{y_m}.}. Also, at each step of modification all dummy indices conflicts will be resolved automatically.

\section{Comparison with existing software}
\label{sec_comparison}
In this section we compare functionality and performance of Redberry with the existing software. At the moment there are a lot of packages and CASs which provide functionality for symbolic tensor manipulation. As far as authors know, the architecture of existing open source systems is based on the principal of {\itshape indices canonicalization} --- an algorithm which brings indices of tensorial expressions to unique canonical order, thereby providing a simple way for expression comparison \cite{RodionovTaranov, Portugal}. For our purposes we choose Cadabra version 1.29  system \cite{Cadabra} which is based on the xPerm \cite{xPerm} library for tensor canonicalization and Mathematica package xAct version 1.1.0 \cite{xAct} (uses the recent xPerm version). Recently, Maple (starting from Maple 18 Physics research version 42.1 released in December 2014) also implemented functionality for simplifying tensorial expressions which were previously unavailable (e.g. most of the examples provided below failed in previous Maple Physics versions). Since Maple and its Physics package are proprietary, the underlying architecture is not known.

Table \ref{table:FeaturesComparison} shows comparison of tensor-specific features between  systems under consideration (both Mathematica and Maple also provides a wide range of "scalar" CAS  features which are not shown in the table). 

\begin{table}[t]
\begingroup
\renewcommand{\arraystretch}{1.5}
\newcommand{\scell}[2][c]{%
  \begin{tabular}[#1]{@{}c@{}}#2\end{tabular}}
\begin{center}
\begin{tabular}{|c|c|c|c|c|}
\hline
Feature & Cadabra &  xAct & Maple Physics & Redberry \\[3pt]
\hline
Multiple index types & Yes &  Yes & \scell{Yes\\[-7pt](4 types available)}& 
\scell{Yes\\[-7pt](8 types available)} \\
Permutational symmetries  & Yes\textsuperscript{1} &  Yes & Yes\textsuperscript{1}%
/No\textsuperscript{2} 
& Yes \\
Multi-term symmetries  & Yes &  Yes & No & No \\
Simple tensor substitutions & Yes & Yes & Yes & Yes \\
\scell{Complicated substitutions\textsuperscript{3}\\[-7pt] (sums, products etc.)} 
& Yes/No\textsuperscript{4} &  No & No & Yes \\
Tensor functions\textsuperscript{5} & No & No &   No & Yes \\
Noncommutative products & Yes &  Yes & Yes & \scell{No/Yes\\[-7pt] (only for 
matrices/spinors)} \\
 Component calculus & No & Yes & Yes & No\\
 Riemann geometry & Yes & Yes & Yes & No\\
\scell{Automatic reduction to standard form /\\[-7pt] automatic 
canonicalization\textsuperscript{6}} & No & No & Yes & Yes \\
 Programming language & No\textsuperscript{7} &  Yes &  Yes & Yes \\ 
 Open source & Yes & No/Yes\textsuperscript{8} &  No & Yes \\
\hline
\end{tabular}
\end{center}
\caption{{\bfseries Comparison of tensor-specific features of selected systems.} 
(1) Cadabra and Maple does not allow to set symmetries equal to arbitrary permutation group, while Redberry allows.
(2) Although Maple allows to set up permutational symmetries of tensors, it allows to simplify only the simples expressions involving symmetries and fails on complicated examples.
(3) Complicated substitution here means a substitution where l.h.s. is a complicated tensor which contains products, sums, tensor fields etc. (see Sec.~\ref{sec_Substitutions}). 
(4) For example, it is impossible to apply substitutions like $K_a (A^{ab} - A^{ba}) = F^b$ in the expression $K_p (A^{qp} - A^{pq})$ (see this example in Sec.~\ref{sec_Substitutions}).
(5) Tensor function represents tensorial function with tensorial arguments (as e.g. {\normalsize\ttfamily\color{listings_base} f\_\{mn\}[x\_\{ab\}, y\_c]} in Redberry) which can be used in substitutions etc. 
(6) For details see Sec.~\ref{sec_StandardForm}. 
(7) Cadabra provides C++ API which in authors opinion is useful for developers but not for users.
(8) xAct is free and open-source, while Mathematica is not.}
\label{table:FeaturesComparison}
\endgroup
\end{table}

\subsection{Performance}
\label{sec_comparison_performance}
In the performance evaluation we focused specifically on manipulation with abstract tensorial expressions avoiding special cases for which different systems can be additionally optimized (like manipulation with Riemann tensor). In order to analyse performance of systems for particular class of problems we proceed as follows. We used a set of randomly generated tensors with different parameters (number of free indices, depth, average product/sum size, set of basic tensors from which the expression was composed). Each initial expression was simplified by expanding out products of sums and reducing similar terms, then rewritten in the equivalent form by: (1) shuffling product/sum terms, (2) renaming dummies and (3) interchanging indices of simple tensors according to their symmetries. Each benchmark was performed on expression obtained by subtracting rewritten expression from the initial one. We measured time needed for different systems to obtain zero by applying simplification operation on such input. Such simplification frequently occurs in real computations and involves many internal components of CAS. 

\newpage
We used the following simplification procedures:
\begin{itemize}
\item 
In case of Redberry we used the following transformation to obtain zero:
\begin{lstlisting}[frame = tb]
def tr = EliminateMetrics & 'd^i_i = 4'.@\ttt@ &
         Expand[EliminateMetrics & 'd^i_i = 4'.@\ttt@] &
         EliminateMetrics & 'd^i_i = 4'.@\ttt@ &
         EliminateDueSymmetries
\end{lstlisting}

\item 
The following code was used in Cadabra (Cadabra spacetime dimension set to 4)\footnote{we tried a variety of combinations of Cadabra transformations and found that this one works faster and covers largest number of cases.}:
\begin{lstlisting}[frame = tb,escapechar = $]
@sumflatten!!(%):@prodflatten!!(%):
@eliminate_metric!(%):@eliminate_kr!(%):
@distribute!(%):@sumflatten!!(%):@prodflatten!!(%):
@eliminate_metric!(%):@eliminate_kr!(%):
@prodsort!(%):@canonicalise!(%):@prodsort!(%):
@rename_dummies!(%):@prodsort!(%):@collect_terms!(%);
\end{lstlisting}
When a single invocation of this code did not produce zero (this occurred in less than 1\%  of cases), we applied it several times (but not more then ten)\footnote{we analysed each of such cases in more detail and found that they can be solved by changing the above sequence of Cadabra transformations in one run}.

\item 
In case of xAct we used function \inlinecode{ToCanonical[]} which performs almost the same steps as described above.

\item 
Maple defines no special transformations for operations with tensors like collect similar terms or eliminate metric. It provides only one transformation \inlinecode{Simplify}\footnote{we've specifically  contacted Maple Physics developers and they suggested using \inlinecode{Simplify} with \inlinecode{tryhard} option which enables \inlinecode{Simplify} to work with tensors; they also assured us that this option will be default in the future Maple releases}, which is used for any simplification of tensorial expressions. When a single invocation of \inlinecode{Simplify} did not produce zero, we applied it twice (increase of the number of retries does not change the number of correct results). In case of non-zero result, test was considered as failed; such cases are excluded from consideration.

\end{itemize}

The expressions used for performance benchmarking can be found at \href{http://redberry.cc/documentation:benchmarks\#downloads}{Redberry website}. The source code used for expressions generation and benchmark execution is available on request.

\subsubsection{Benchmarking of atomic operation}
In our first benchmark we measured performance of {\itshape comparison operation}, which is probably the most frequent atomic operation arising in any calculation. For this purpose, we generated sums of products of tensors and measured time needed to obtain zero when subtracting it (rewritten in equivalent form as described earlier) from itself.

As we have found, the existing systems fails on some expressions with complicated structure. For example, if tensor $W_{abcde}$ is fully symmetric in indices $a,c,e$ then the following expression is zero:
\begin{multline*}
T^{h} \,
(W_{bde}{}^{ij}+W_{bde}{}^{ji}+W_{bed}{}^{ij}+W_{dbe}{}^{ji}+W_{de}{}^{i}{}_{b}{}^{
j}) (W_{cfhji}+W_{chfji}+W_{cjhfi}+W_{fchij}+W_{fchji})\,\, 
- \\
T^{h} \,
(W_{bde}{}^{ij}+W_{bde}{}^{ji}+W_{bed}{}^{ji}+W_{dbe}{}^{ij}+W_{de}{}^{i}{}_{b}{}^{
j}) (W_{cfhij}+W_{chfij}+W_{cihfj}+W_{fchij}+W_{fchji}) \,=\, 0
\end{multline*}
This result directly follows from the structure of summands and can be obtained by relabelling dummy indices and interchanging indices according to symmetries. Redberry simplifies this expression in approximately 2 milliseconds (right at parse) without any additional calculations. On the other hand, other systems need to expand out product of sums in order to determine that the result is zero; this, of course, is much more expensive calculation. In order to avoid additional expand operation, we generated sums (of size 100) of products of simple tensors with fixed number of free indices. So, the typical input problem looked as follows:
\[
\left( W{}^{\alpha\beta} \, T{}^{\nu}{}_{\beta\rho} \, R{}^{\theta\gamma\rho} \, F{}_{\mu\theta\alpha} \, + \, \dots \mbox{\itshape\small\color{grey} (99 terms)} \, \right) 
\,-\,
\left( F{}_{\mu\rho\theta} \, W{}^{\theta\beta} \, R{}^{\rho\gamma\alpha} \, T{}^{\nu}{}_{\beta\alpha} \,  +\, \dots \mbox{\itshape\small\color{grey} (99 terms)} \, \right) \, = \, 0
\]

\begin{figure}[h]
  \centering
  \begin{tabular}{cc}
    \includegraphics[width=0.5\textwidth]{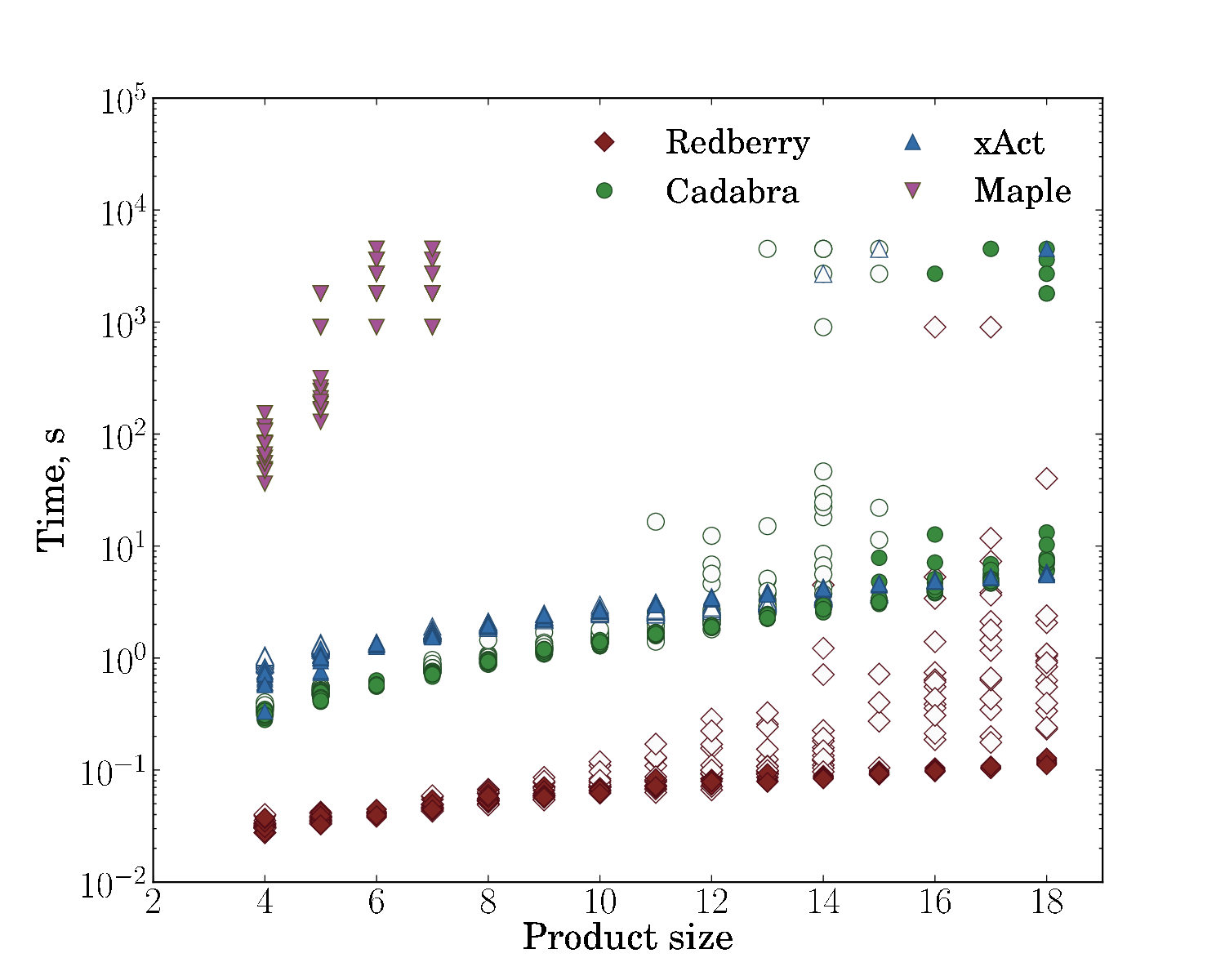} & 
    \includegraphics[width=0.5\textwidth]{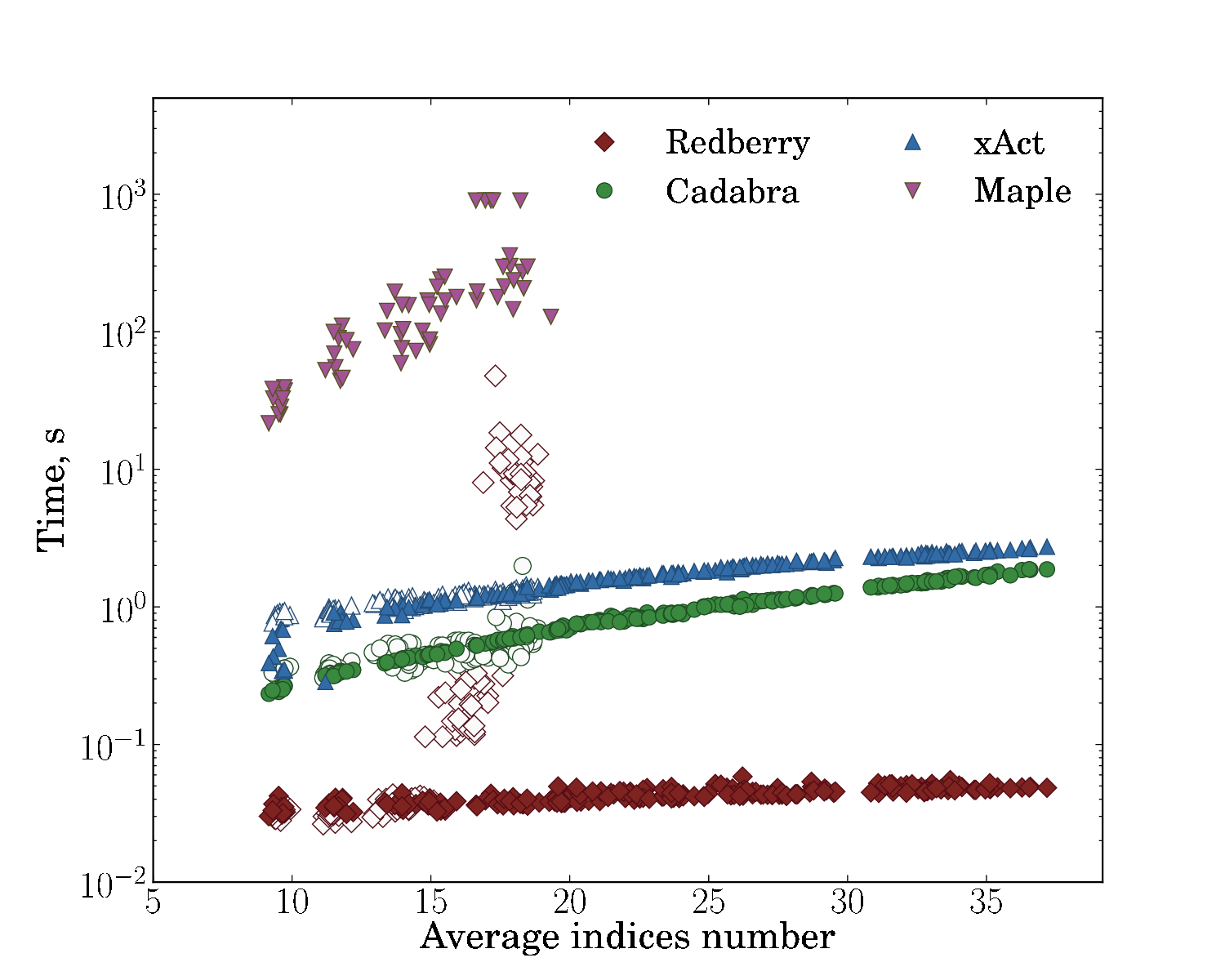} \\
  \end{tabular}
  \caption{\textbf{Dependence of time spent in simplification routine on product size (left) and on average number of indices in products (right)}. For these benchmarks we used sums (100 summands) of products of simple tensors, then subtracted it from itself (in equivalently rewritten form: shuffled summands and multipliers, renamed dummy indices and interchanged indices of simple tensors according to their symmetries) and measured time needed to obtain zero when simplifying inputs prepared in such a way. For Redberry, xAct and Cadabra we performed benchmarks both with (blank markers) and without symmetries (filled markers) of simple tensors, while Maple was unable to simplify such expressions in case with symmetries. The expressions used for these benchmarks can be found at \href{http://redberry.cc/documentation:benchmarks\#downloads}{Redberry website}.}
\label{fig:metrics}
\end{figure}

The results of our benchmarking\footnote{all benchmarks were performed on the same hardware: AMD Phenom II X6 1100T; 16 Gb RAM; Ubuntu Linux 12.04} are presented on Fig.~\ref{fig:metrics}. Left figure shows the dependence of time needed by system to obtain zero on the number of multipliers in products in target sum. In this benchmark we used a set of basic tensors with number of indices from 1 to 6 (e.g. $A_a$, $B_{ab}$, $C_{abc}$ etc.), number of ``outer'' free indices equal to 5 and varying number of product lengths from 4 to 18 (increasing this parameter, we thereby increased number of dummy indices).

We analysed both situation when simple tensors have symmetries (symmetric or antisymmetric with respect to all its indices) and not; unfortunately, in the first case Maple was unable to obtain zero on almost all used examples, so we excluded Maple in this case. In case without symmetries Redberry shows polynomial behaviour on number of multipliers, while Cadabra and xAct have small exponent and Maple has extremely large exponent or even factorial dependence (which shows that it uses some brute-force algorithm). In case with symmetries, all three systems (Cadabra, xAct and Redberry) have exponential dependence in the worst case, but Redberry has better behaviour on larger problems.

For the benchmark presented on the right picture in Fig.~\ref{fig:metrics}, we used sums of products with fixed length equal to 5, fixed number of ``outer'' free indices equal to 5 and with varying number of basic simple tensors with different number of indices (thereby affecting number of dummy indices) and measured time dependence on average number of indices in products. As in the first benchmark in the case without symmetries we observed polynomial dependence for Redberry, small exponent for xAct and Cadabra, and factorial in case of Maple. On the other hand,  presence of symmetries almost does not change the behaviour of Cadabra and xAct, while Redberry failed to simplify large expressions in a reasonable time.

\begin{figure}[h!]
  \centering
  \begin{tabular}{cc}
    \includegraphics[width=0.5\textwidth]{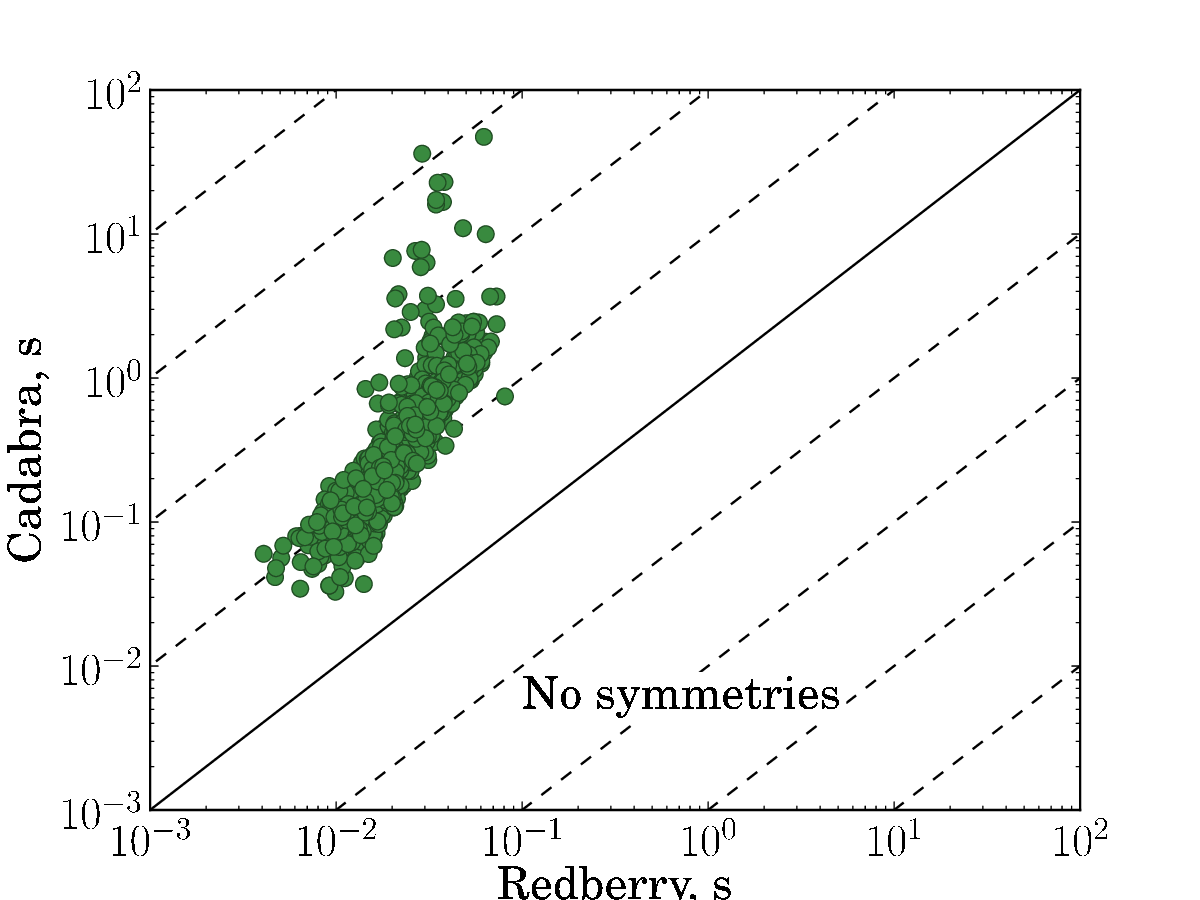} &
     \includegraphics[width=0.5\textwidth]{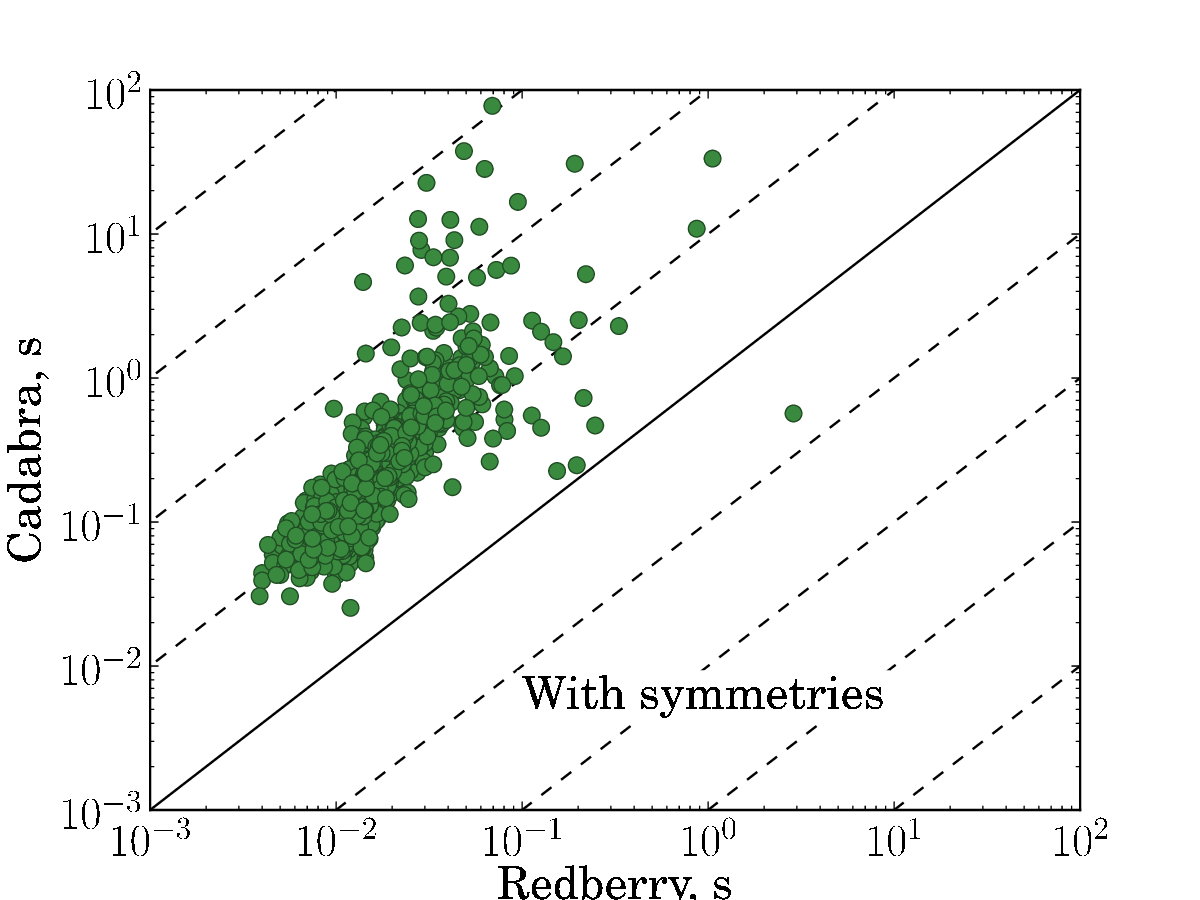}\\
     \includegraphics[width=0.5\textwidth]{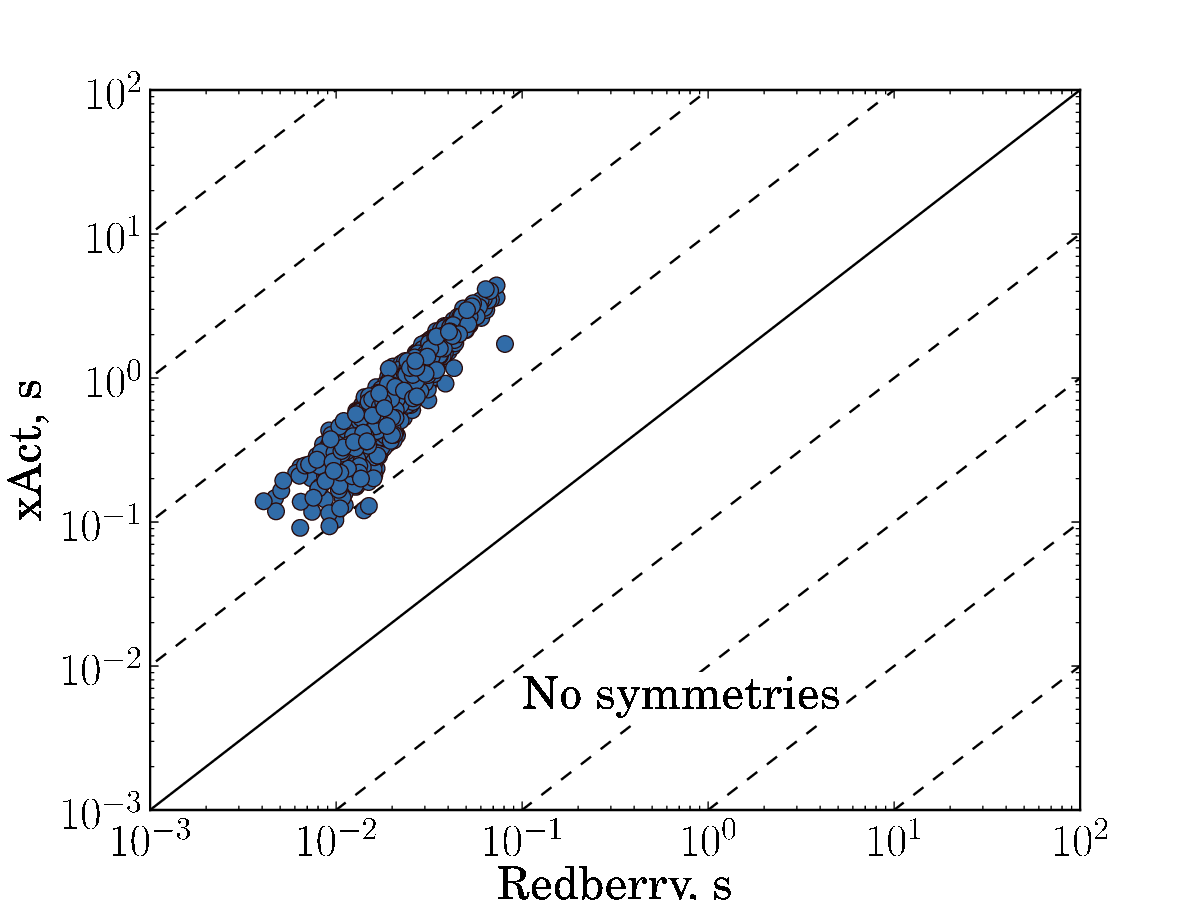}  &
     \includegraphics[width=0.5\textwidth]{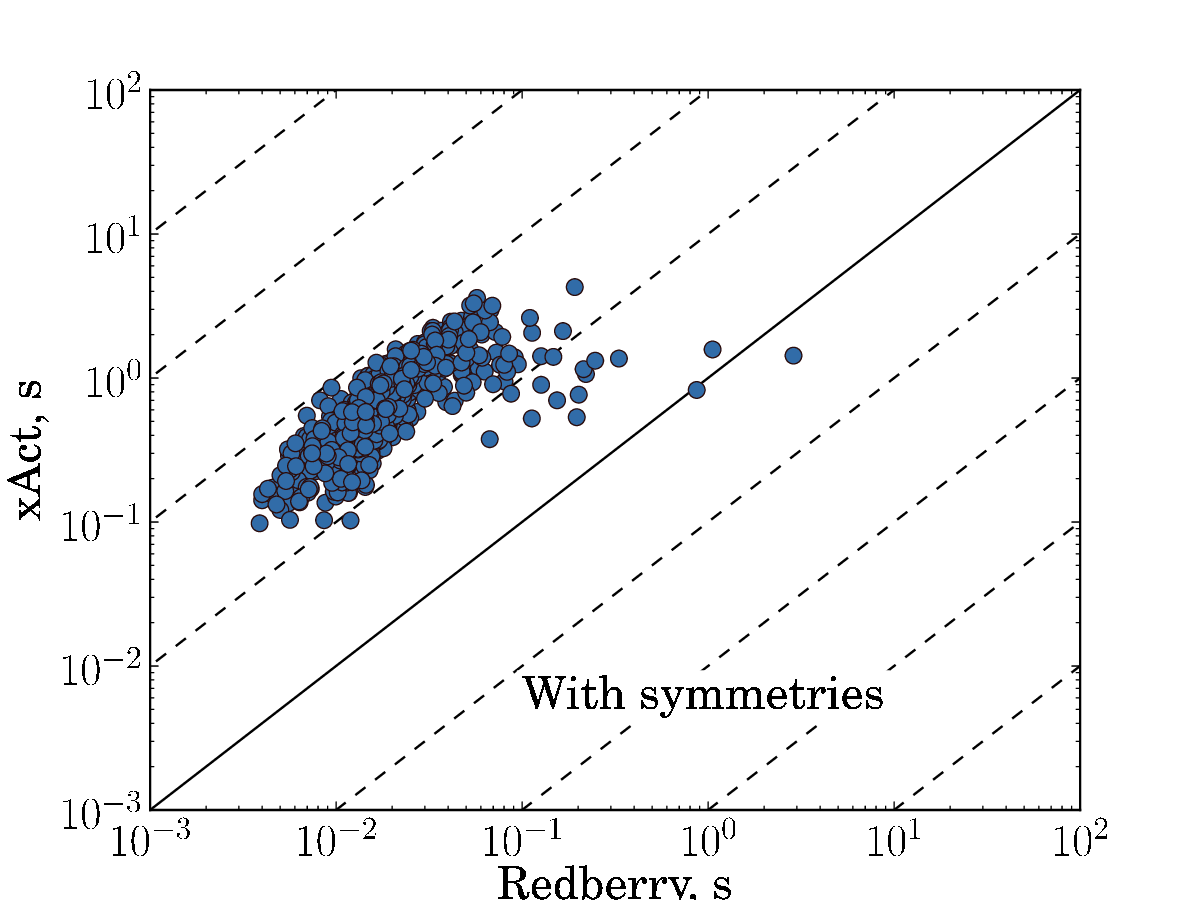} \\  
     \includegraphics[width=0.5\textwidth]{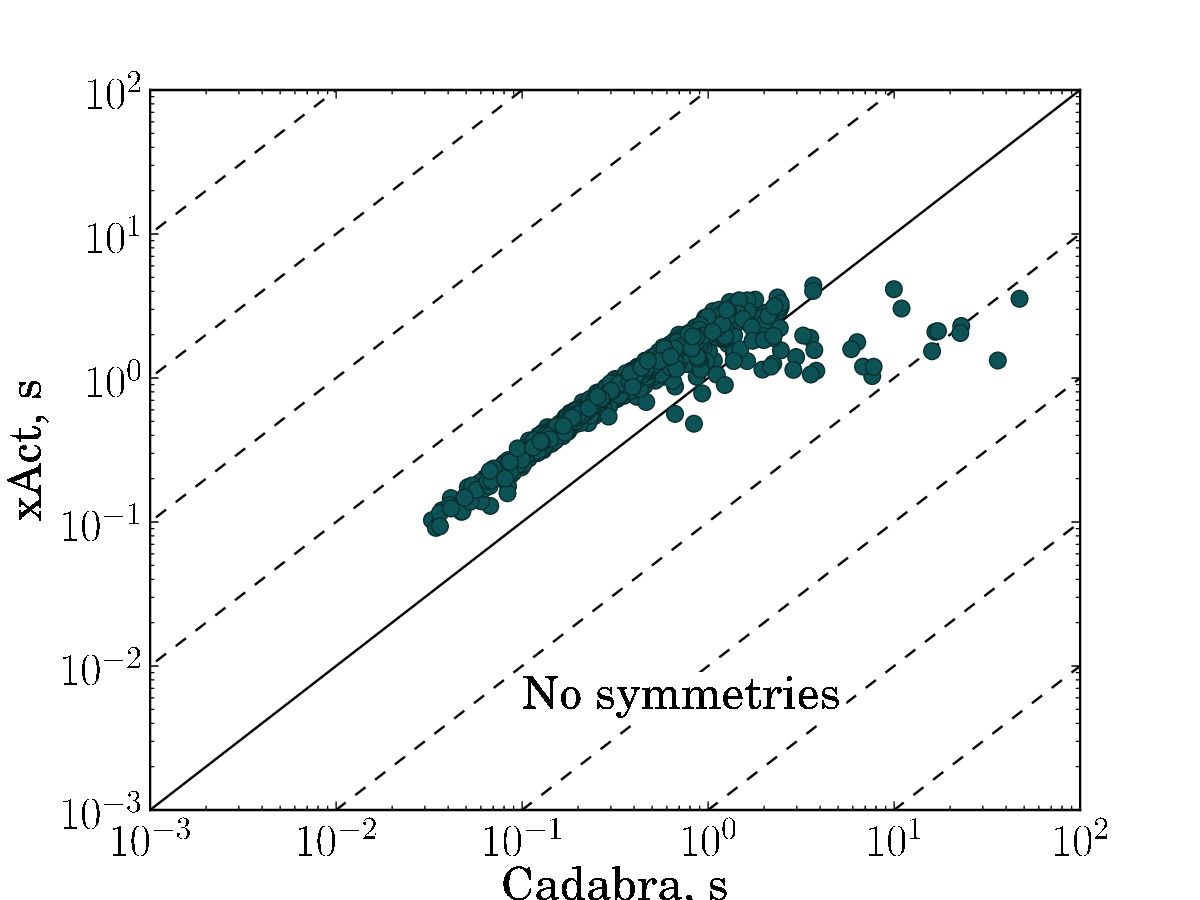} &
      \includegraphics[width=0.5\textwidth]{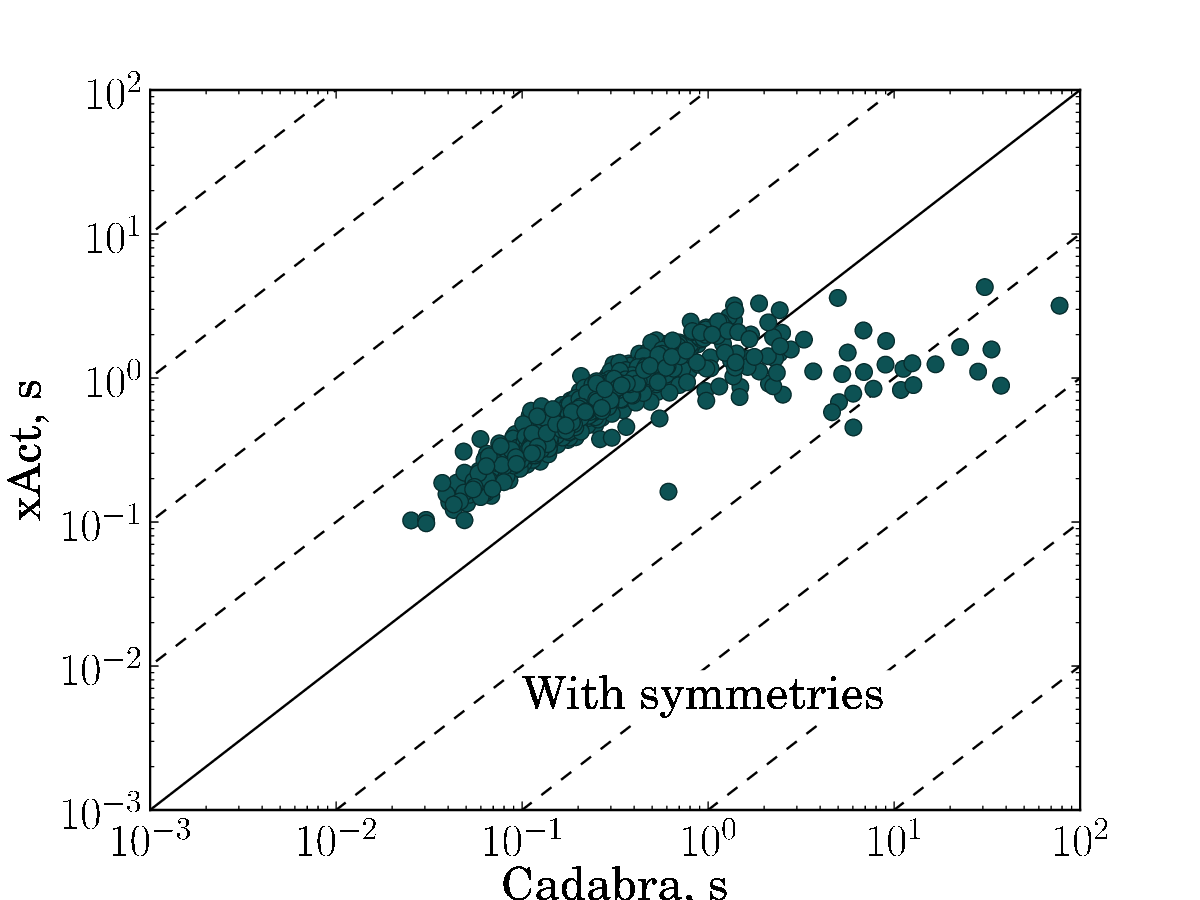}
  \end{tabular}
  \caption{\textbf{Comparison of time spent in simplification routine for randomly generated expressions.} Each input problem plotted as filled circle. Each plot represents comparison of execution times needed to simplify input expression to zero between different systems (xAct versus Redberry, Cadabra versus Redberry, xAct versus Cadabra). Solid line corresponds to equal execution times, dashed lines shows 10x, 100x, 1000x, etc. execution time ratio. On average in presented benchmarks Redberry is 41 times faster than xAct and 29 times faster than Cadabra in case without symmetries and 26 and 33 times faster in case with symmetries. The expressions used for these benchmarks can be found at \href{http://redberry.cc/documentation:benchmarks\#downloads}{Redberry website}.}
\label{fig:simplify}
\end{figure}

\subsubsection{Benchmarking of simplification routine}

Another important metric of CAS is its performance in simplification of expressions with deeply nested structure. For this benchmarks we generated sums of products of sums of products (so expression depth is 4) with following characteristics: we used basic tensors $A_a$, $B_{ab}$, $C_{abc}$, all generated expressions had one free index, varying product sizes from 2 to 4 and varying sum sizes from 2 to 4. So the used expressions looked like:

\[
\left( F_{\mu\nu} \left( T^{\mu\alpha} R^{\nu\beta} \,+\, \dots \right) \,+\, \dots \right)
\left( R_{\alpha\rho} \left( F^{\rho}{}_{\beta} R_{\tau\gamma}\phantom{{}^\beta} +\, \dots \right) \,+\, \dots \right) \,\times\, \left( \phantom{{}^\beta}  \dots \phantom{{}^\beta} \right)
 \,- \, 
\left( F_{\nu\mu}T^{\nu\beta}R^{\mu\alpha}R_{\beta\rho}F^{\rho}{}_{\alpha} R_{\tau\gamma}  \,+\, \dots \right) \,=\, 0
\]

Unfortunately, Maple Physics was unable to simplify any of provided examples\footnote{we've contacted with Maple Physics developers; they answered that it is unknown when functionality for simplification of these examples will be available}, so we had to exclude Maple from consideration. 

Fig.~\ref{fig:simplify} shows the comparison of time spent in {\itshape expand and collect} operation for the same input expressions for Redberry vs. xAct, Redberry vs. Cadabra and xAct vs. Cadabra. We performed tests both in case when tensors have symmetries ($B_{ab}$ antisymmetric and $C_{abc}$ symmetric) and not. Summarizing all timings and taking the average, we found that Redberry is 41 times faster than xAct and 29 times faster than Cadabra in case without symmetries and 26 and 33 times faster in case with symmetries in our benchmarks. It is worth noting that such a performance gain can become much more significant in case of more complicated expressions, where size of intermediate expressions became larger, and in such cases Redberry can be even several thousands times faster because of its ability to simplify intermediate expressions during evaluation.

\section{Conclusions}
In this paper we presented Redberry --- an open source computer algebra system designed to manipulate with symbolic tensorial expressions. Redberry is a computer algebra system which considers both tensors and indexless expressions in a common way. It provides basic tensor-specific features such as tensor symmetries, multiple index types, dummy indices handling, \LaTeX-style I/O, mappings of tensor indices and a comprehensive set of tensor-specific transformations etc. Rich functionality of Redberry was illustrated on a complex physical problems: Feynman graphs and one-loop counterterms calculations. Redberry provides a simple and convenient facade to modern high-level programming language (Groovy), which makes it possible to use all features of general-purpose programming language combined with domain-specific features provided by the CAS.

Throughout this paper we illustrated main Redberry features with many examples. In Sec.~\ref{sec_Basics} we described the basic usage of Redberry, how it handles mathematical expressions and how different properties of expressions can be accessed through Groovy syntax. In Sec.~\ref{sec_BasicTransformations} we gave a list of the most common transformations and showed how they	 can be manipulated in Redberry. In Sec.~\ref{sec_PhysExamples} we illustrated functionality of Redberry on two real world examples: calculation of Feynman graphs and calculation of one-loop counterterms in curved space-time. We also illustrated handling of matrix objects  (such as spinors or Dirac matrices) using convenient syntax in Redberry. Sec.~\ref{sec_InternalArchitecture} was dedicated to the quintessential internals of Redberry --- mappings of indices and expression-tree traversal. In particular, in this section we discussed how a graph-theoretical approach to tensor manipulation is used in Redberry and how it compares with approaches used by other tensor-oriented CASs. Finally, in Sec.~\ref{sec_comparison} we gave an overview of Redberry features and performance in comparison with other well-known CASs.

In this paper we showed that a new approach to tensor manipulation based on graph-theoretical algorithms allows to achieve considerably better performance on a large set of typical algebraic problems than other systems mentioned in Sec.~\ref{sec_comparison}. Such a high performance enables Redberry to solve a large-scale real-world problems in high energy physics which requires manipulation with huge tensorial expressions. Still, there are many important features which are to be implemented in Redberry: complete pattern matching, multi-term symmetries, Riemann geometry, noncommutativity, ``scalar'' transformations etc.

Redberry is licensed under GNU GPLv3 and anyone can contribute in development of Redberry. Comprehensive documentation, examples and installation instructions are available at \href{http://redberry.cc}{http://redberry.cc}. Known issues and release schedule can be found at \href{http://youtrack.redberry.cc/issues}{http://youtrack.redberry.cc}. The source repository can be found at \href{http://bitbucket.org/redberry/redberry}{http://bitbucket.org/redberry}.

\section*{Acknowledgments}
The authors are grateful to Petr Pronin and Konstantin Stepanyantz for a useful discussion. The work of S.P. was financially supported by Russia Foundation for Basic Research (grants \#12-02-31249a, \#12-02-31408a and \#14-02-00096 А), grant of the president of Russian Federation (grant \#MK-3513.2012.2) and grant of SAEC "Rosatom" and  Helmholtz Association. The work of D.B. was financially supported by the Molecular and Cell Biology program of the Russian Academy of Sciences.


\end{document}